\documentclass[pre,onecolumn,preprintnumbers,floatfix,amssymb,amsmath]{revtex4}
\usepackage{graphicx}
\usepackage{natbib}
\usepackage{subfigure}
\usepackage{color}
\usepackage{multirow}
\usepackage{amssymb}
\usepackage{subfigure}
\usepackage{amsmath}

\graphicspath{}

\begin{document}

\title{Buoyancy and capillary effects on floating liquid  lenses}
\author{P. D. Ravazzoli$^1$, A. G. Gonz{\'a}lez$^1$, J. A. Diez$^1$, H. A. Stone$^2$}
\affiliation{$^1$ Instituto de F\'{\i}sica Arroyo Seco, Universidad Nacional del Centro de la Provincia de Buenos Aires, and CIFICEN-CONICET-CICPBA, Pinto 399, 7000, Tandil, Argentina\\
$^2$Department of Mechanical and Aerospace Engineering, Princeton University, Princeton, NJ 08544 USA}

\begin{abstract}
We study the equilibrium shape of a liquid drop resting on top of a liquid surface, i.e., a floating lens. We consider the surface tension forces in non--wetting situations (negative spreading factor), as well as the effects of gravity. We obtain analytical expressions for the drop shape when gravity can be neglected. 
Perhaps surprisingly, when including gravity in the analysis, we find two different families of equilibrium solutions for the same set of physical parameters. These solutions correspond to drops whose center of mass is above or below the level of the external liquid surface. By means of energetic considerations we determine the family that has the smallest energy, and therefore is the most probable to be found in nature. A detailed explanation of the geometrical differences between the both types of solutions is provided.
\end{abstract}

\maketitle

\section{Introduction}

Fluid--fluid interactions between two immiscible liquids are common in nature and in many industrial processes. Pioneering work goes back at least to Benjamin Franklin~\cite{Franklin1774}, but later on a plethora of papers have been devoted to the spreading phenomenon of one liquid over another (see e.g. Lord Rayleigh~\cite{Rayleigh1890}, Neumann and Wangerin~\cite{Neumann1894}, Hardy~\cite{Hardy1912}, Lyons~\cite{Lyons1930}, Langmuir~\cite{Langmuir1933}, Miller~\cite{Miller1941}, Zisman~\cite{Zisman1941}, Seeto et al.~\cite{Seeto1983}, Takamura et al.~\cite{Takamura2012}).

In more recent years, the wettability of liquids over liquids has continued to be studied with focus on new features. For example, Wyart~et~al.~\cite{Wyart1993} studied liquid films dewetting from another liquid. Burton~et~al.~\cite{Burton2010}, and more recently Tress~et~al.\cite{Tress2017}, analyzed the shape of a liquid lens, while Chen~et~al.~\cite{Chen1997} studied the dependence of the lens size on the contact angle, and McBride~et~al.~\cite{McBride1988}, Endoh~et~al.~\cite{Endoh1990}, Levich~et~al.~\cite{Levich1969} and Sebilleau~et~al.~\cite{Sebilleau2018} have been concerned with the spreading phenomenon.

Physically, a liquid lens is a drop lying over another immiscible liquid and surrounded by a gas phase,  such as air. At the equilibrium, the three phases meet along a circular line, where the sum of the three tensions must be zero. Neumann's rule~\cite{Nikolov2017, Iqbal2017} is the corresponding version of the Young equation for a solid substrate and provides the balance between the tensions at the contact line.

Recently, several authors have addressed different aspects of floating lenses. The evaporation process of a liquid lens has been considered in~\cite{Lu2019} and compared with a theoretical model constructed assuming a constant contact angle and spherical cap shape. Also, the interaction, coalescence and repulsion of floating drops was studied in \cite{He2019} and \cite{Vinay2019}.

To study the dynamic behavior of a liquid lens, it is necessary to get a full understanding of the static case and how the physical parameters affect the shape of the resulting drop. We present here some aspects of the static solutions that have not been fully addressed. The liquid lens shape has been studied previously by several authors~\cite{Ross1992, Burton2010, Clint2017}. In the present work, we initially follow the guidelines presented by Burton~et~al.~\cite{Burton2010} for the partial wetting situation, and numerically solve the three pairs of coupled differential equations resulting from the pressure balance on each surface. Firstly, we present the basic formalism in Sec.~\ref{sec:TheoSol} and define the appropriate dimensionless parameters. We characterize the problem by using three parameters, namely, a reference Bond number, $Bo$, the ratios of surface tensions, and the dimensionless spreading factor. Then, we analytically solve the problem without considering gravity effects, and obtain expressions for the two spherical caps that constitute an equilibrium floating drop.
In Sec.~\ref{sec:GravitySolutions} we take into account the gravity effects and identify the existence of two families of solutions for the same set of parameters. To the best of our knowledge, this interesting result has not been reported previously. Here, we show its existence and give a detailed description. In order to decide which type of solutions is more likely to be found in nature, we perform an energetic analysis in Sec.~\ref{sec:energy}. We calculate the system energy for the different scenarios and find that one family of solutions is always lower energy than the other. 

\section{Description of the problem and formalism}\label{sec:TheoSol}
\subsection{Governing equations and boundary conditions}

\begin{figure}[t]
\includegraphics[width=0.7\linewidth]{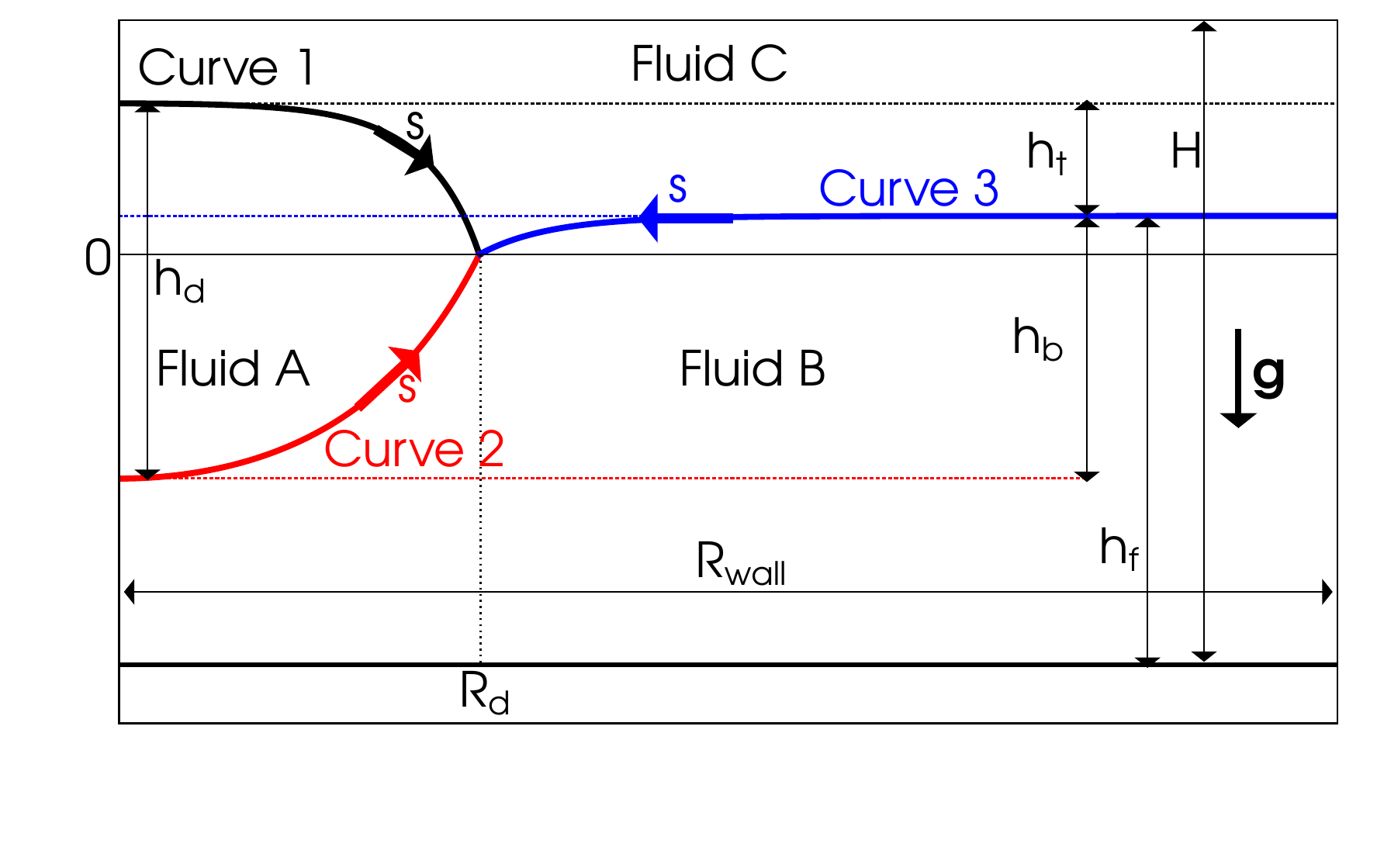}
\caption{Dimensionless scheme of a liquid lens (fluid A) over a deep external liquid (fluid B), surrounded by air (fluid C). The triple contact line is at $r=R_d$ and $z=0$ ($R_d \ll R_{wall}$). The length scale is ${\cal R}_0$ (Eq.~\ref{eq:R0}).}
\label{fig:scheme}
\end{figure}

We are interested in the shapes of static interfaces that develop when a drop is deposited on a liquid surface under the effects of both surface tension and gravity. In particular, we consider the case when a drop (fluid A) floats under partial wetting conditions on the liquid--air interface (fluids B, C); see Fig.~\ref{fig:scheme}. To scale the problem, we use a characteristic length scale given by
\begin{equation}
{\cal R}_0= \left( \frac{3 {\cal V}_0}{4 \pi} \right)^{1/3},
\label{eq:R0}
\end{equation}
where ${{\cal V}_0}$ is the volume of the drop. 

In this dimensionless configuration (see Fig.~\ref{fig:scheme}), we assume that the drop radius, $R_d$, is much smaller than the size of the container (of radius $R_{wall}$), and that the thickness of the lower layer, $h_f$, is always large enough to assure that the drop never touches the bottom no matter the drop volume, $V={{\cal V}_0}/{\cal R}_0^3$. The interface curves between each fluid are denoted by numbers. Thus, curve $1$ corresponds to the interface between fluids A and C, curve $2$ to A and B, and curve $3$ to B and C. As shown in Fig.~\ref{fig:scheme}, the arc length along each curve, $s$, increases from $0$ towards the triple point, where the three curves meet. 

Since surface tension forces are responsible for the Laplace pressure jump across the liquid interfaces, we can write
\begin{equation}
p_{i} - p_{j}=\sigma \kappa,
\end{equation}
where $p_i$ and $p_j$ are the hydrostatic pressures in the bulk of each fluid at both sides of the corresponding curve of curvature $\kappa$, so that $i$ and $j$ stand for A, B or C. Thus, the equilibrium equation for each interface can be written
\begin{subequations}
\label{eq:equil_dim}
\begin{equation}
(P_C - P_A) + g z (\rho_A - \rho_C) = \sigma_{1} \kappa_{1},
\end{equation}
\begin{equation}
(P_A - P_B) + g z (\rho_B - \rho_A) = \sigma_{2} \kappa_{2},
\end{equation}
\begin{equation}
(P_B - P_C) + g z (\rho_C - \rho_B) = \sigma_{3} \kappa_{3},
\end{equation}
\end{subequations}
where $P$ refers to the reference pressure inside each fluid and, the subscripts in capital letters and numbers, respectively, correspond to a fluid and the interfaces between them. The \emph{dimensionless} form of these equations can be written as
\begin{equation}
\Delta P_i + Bo_i \, z_i(s) = \kappa_i(s), \qquad i=1,2,3,
\label{eq:equil_adim}
\end{equation}
where all lengths are expressed in units of ${\cal R}_0$, and we have defined the \emph{dimensionless} constants
\begin{subequations}
\begin{equation}
\Delta P_1= \frac{{\cal R}_0}{\sigma_1} (P_C - P_A),
\end{equation}
\begin{equation}
\Delta P_2 = \frac{{\cal R}_0}{\sigma_2} (P_A - P_B),
\end{equation}
\begin{equation}
\Delta P_3 =  \frac{{\cal R}_0}{\sigma_3} (P_B - P_C),
\end{equation}
\begin{equation}
Bo_1 = \left( \frac {\rho_A - \rho_C}{\sigma_1}\right) g {\cal R}_0^2,
\end{equation}
\begin{equation}
Bo_2  = \left( \frac {\rho_B - \rho_A}{\sigma_2}\right) g {\cal R}_0^2,
\end{equation}
\begin{equation}
Bo_3  = \left( \frac {\rho_C - \rho_B}{\sigma_3} \right) g {\cal R}_0^2.
\end{equation}
\end{subequations}

Assuming axial symmetry for this problem, the \emph{dimensionless} surface curvature is given by~\cite{Burton2010}
\begin{equation}
\kappa_i = \frac{z_i'(s)}{r_i(s) [r_i'(s)^2+z_i'(s)^2]^{1/2} } +
\frac{r_i'(s) z_i''(s)-z_i'(s) r_i''(s)}{[r_i'(s)^2+z_i'(s)^2]^{3/2}},
\end{equation}
where $'$ denotes the derivative with respect to $s$. If $L_i$ is the entire arc length of any of the curves ($1$, $2$ or $3$), then we scale the arc length as $q=s/L_i$ ($0\leq q \leq 1$) so that
\begin{equation}
\kappa_i = \frac{z_i'(q)}{r_i(q) L_i } + \frac{r_i'(q) z_i''(q)-z_i'(q) r_i''(q)}{L_i^3},
\end{equation}
where 
\begin{equation}
L_i^2=r_i'(q)^2+z_i'(q)^2=\text{const}. 
\label{eq:Ldef}
\end{equation}
This condition allows us to obtain two equations for $(r_i(q),z_i(q))$ as:
\begin{subequations}
\begin{equation}
r_i''(q) =  z_i'(q) \left[ \frac{z_i'(q)}{r_i(q)} - L_i \kappa_i \right],
\end{equation}
\begin{equation}
z_i''(q) = -r_i'(q) \left[ \frac{z_i'(q)}{r_i(q)} - L_i \kappa_i \right].
\end{equation}
\label{eq:r2z2}
\end{subequations}
By replacing here the curvatures from Eq.~(\ref{eq:equil_adim}), we obtain the three pairs of equations for $(r_i(q),z_i(q))$ along each interface ($i=1,2,3$),
\begin{subequations}
\begin{equation}
r_i''(q) = \,z_i'(q) \left[ \frac{z_i'(q)}{r_i(q)} - L_i \Delta P_i - L_i Bo_i \, z_i(s) \right],
\end{equation}
\begin{equation}
z_i''(q) = -r_i'(q) \left[ \frac{z_i'(q)}{r_i(q)} - L_i \Delta P_i - L_i Bo_i \, z_i(s) \right].
\end{equation}
\label{eq:rizi}
\end{subequations}

The integration of all three curves starts with zero slopes at the corresponding $q=s/L=0$, and ends at the contact point. At the beginning point ($q=0$) we have $z_1'=z_2'=z_3'=0$, and according to Eq.~(\ref{eq:Ldef}) it must be $r_1'=L_1$, $r_2'=L_2$ and $r_3'=-L_3$, since $r$ increases along curves $1$ and $2$, and decreases along curve $3$.
The three curves meet at the triple contact point at $q=1$, where $r_i(1)=R_d$ and $z_i(1)=0$. The condition on $z$ is arbitrary since the system is translationally invariant in this direction, since the gravitational potential is linear in $z$. Thus, we will use this property to start all three integrations from $z=0$, and proceed to make the corresponding vertical displacements {\it a posteriori}.

\begin{figure}[htb]
\includegraphics[width=0.6\linewidth]{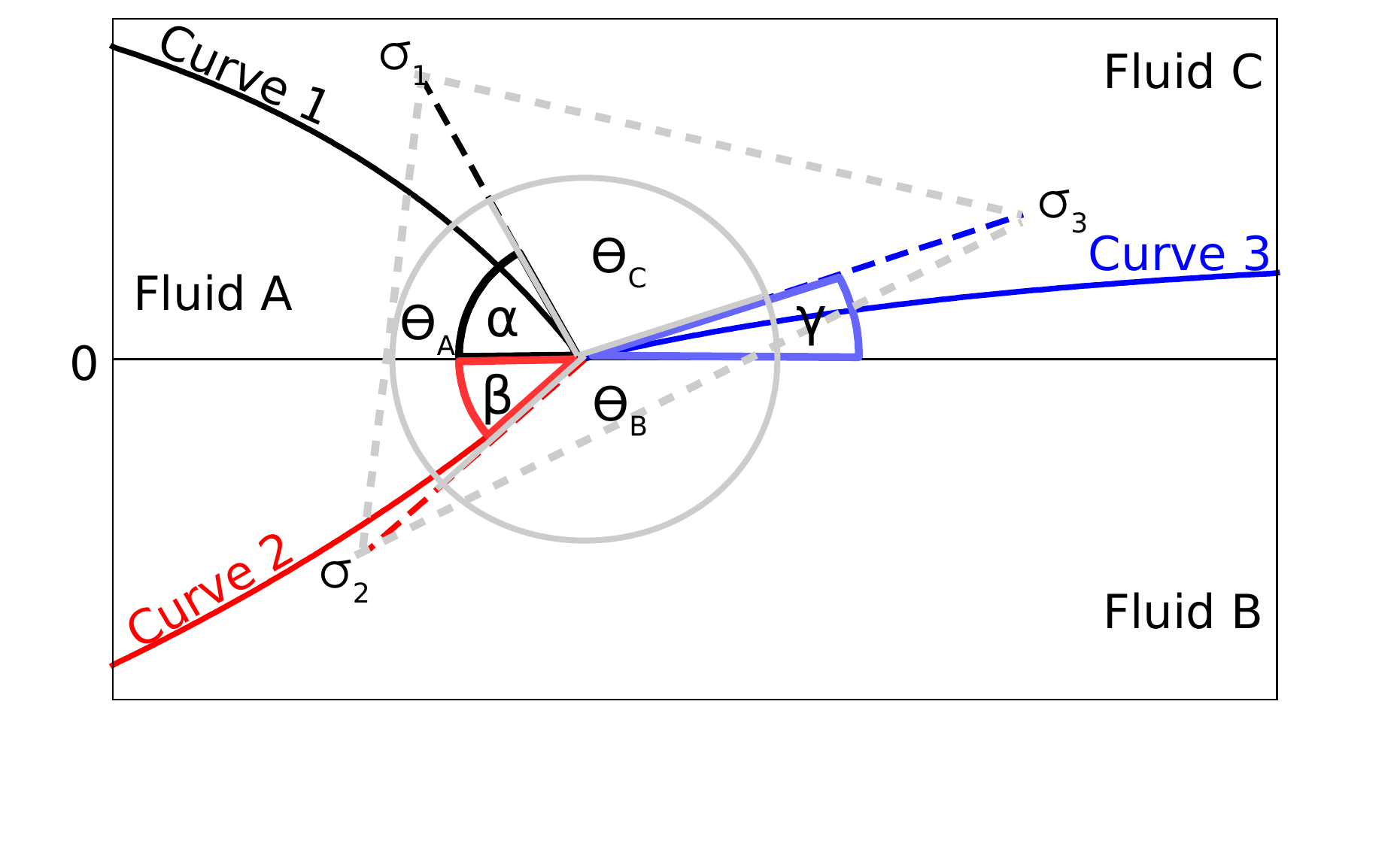}
\caption{Contact angle definitions for a liquid lens (fluid A) over a deep liquid substrate (fluid B), surrounded by air (fluid C).}
\label{fig:angles_scheme}
\end{figure}

Note that both $L_i$ and $\Delta P_i$ are not known {\it a priori} in Eqs.~(\ref{eq:rizi}). These six constants and the drop radius, $R_d$, must be determined consistently by solving all six equations plus the conservation of drop volume. The first three conditions are
\begin{equation}
r_1=r_2 =r_3=R_d,
\label{eq:rq1}
\end{equation}  
at $q=1$. The fourth and fifth constraints are related to the so--called Neumann's rule, i.e., that surface tension forces must equilibrate along both $r$ and $z$-directions (see Fig.~\ref{fig:angles_scheme}),
\begin{subequations}
\label{eq:Neumann}
\begin{equation}
\sigma_1 \cos \alpha + \sigma_2 \cos  \beta - \sigma_3 \cos \gamma =0,
\end{equation}
\begin{equation}
\sigma_1 \sin \alpha - \sigma_2 \sin  \beta + \sigma_3 \sin \gamma =0,
\end{equation}
\end{subequations}
at $q=1$.
The sixth condition is concerned with the evaluation of the reference pressures, $P_i$. In fact, by summing up the three equations in Eq.~(\ref{eq:equil_dim}) at $z=0$, 
\begin{equation}
\sigma_1 \kappa_1 + \sigma_2 \kappa_2 + \sigma_3 \kappa_3 =0,
\label{eq:sum_curv}
\end{equation}
at $q=1$. Finally, the seventh condition is of the integral type, since it refers to the constraint of a given drop volume. Thus, we have:
\begin{equation}
V = \frac{{{\cal V}_0}}{{\cal R}_0^3} = \int_0^1 2 \pi r_1(q) r'_1(q) z_1(q) dq - \int_0^1 2 \pi r_2(q) r'_2(q) z_2(q) dq= \frac{4 \pi}{3}.
\label{eq:lens_vol}
\end{equation}
Therefore, the seven conditions in Eqs. (\ref{eq:rq1})-(\ref{eq:lens_vol}) determine the shapes of the interfaces as well as the values of $L_i$, $\Delta P_i$ and $R_d$. 

A detailed analysis of the Neumann equilibrium conditions allows to find relationships between the contact angles and the surface tensions. In fact, by using the cosine law in the triangle depicted in Fig.~\ref{fig:angles_scheme}, we have
\begin{subequations}
\begin{equation}
\cos \theta_A = \frac{\sigma_3^2-\sigma_2^2-\sigma_1^2}{2 \sigma_1 \sigma_2},\label{eq:cos_lawA}
\end{equation}
\begin{equation}
\cos \theta_B = \frac{\sigma_1^2-\sigma_2^2-\sigma_3^2}{2 \sigma_2 \sigma_3},\label{eq:cos_lawB}
\end{equation}
\begin{equation}
\cos \theta_C = \frac{\sigma_2^2-\sigma_1^2-\sigma_3^2}{2 \sigma_1 \sigma_3},\label{eq:cos_lawC}
\end{equation}
\label{eq:cos_law}
\end{subequations}
which lead to restrictions on the admissible values of the \emph{spreading coefficient},
\begin{equation}
S=\sigma_3-\sigma_2-\sigma_1,
\label{eq:S}
\end{equation}
which is also used to describe the contact line motion on solid substrates. For convenience, we also write Eqs.~(\ref{eq:cos_law}) in terms of contact angles as (see Fig.~\ref{fig:angles_scheme})
\begin{equation}
\alpha + \beta =  \arccos \left( \frac{\sigma_3^2 - \sigma_2^2 - \sigma_1^2}{2 \sigma_1 \sigma_2}\right), \qquad
\alpha + \gamma = \pi - \arccos \left( \frac{\sigma_2^2 - \sigma_1^2 - \sigma_3^2}{2 \sigma_1 \sigma_3}\right).
\label{eq:NewmanRelations}
\end{equation}
Note that the knowledge of one angle and the three interfacial tensions automatically determines the other two.

From the fact that the moduli of Eqs.~(\ref{eq:cos_law}) must be less than one, we find the following restrictions for the spreading coefficient:
\begin{equation}
S < 0,\qquad -2 \sigma_2 < S, \qquad -2 \sigma_1 < S,
\end{equation}
which can be summarized as
\begin{equation}
-2 \min(\sigma_1 , \sigma_2)<S<0.
\label{eq:SCond}
\end{equation}
Therefore, the condition for partial wetting (i.e., the formation of a static floating drop) is more restrictive than in the case of partial wetting of a solid substrate, which simply requires $S<0$. 

\subsection{Nondimensionalization}

Since there are many dimensional parameters necessary to determine the final equilibrium solution, it is useful to define the problem in terms of a fewer number of dimensionless variables. In order to develop this description, it is necessary to select reference values for both density and surface tension, namely,  $\rho_{\text{ref}}$ and $\sigma_{\text{ref}}$, respectively. For convenience we choose these values as $\rho_{\text{ref}} = \rho_A$ and $\sigma_{\text{ref}} = \min (\sigma_1 , \sigma_2)$, where the latter selection is suggested by the condition in Eq.~(\ref{eq:SCond}). 

If we define the ratio 
\begin{equation}
\zeta= \frac{S}{S^\ast}, \qquad S^\ast=-2\sigma_{\text{ref}},
\label{eq:zetaS}
\end{equation}
where $S^\ast$ is a the reference spreading coefficient, all possible solutions correspond to the interval $0\leq  \zeta \leq 1$. So, from Eq.~(\ref{eq:S}) we can write
\begin{equation}
\zeta = \frac{1}{2} \left(\eta_1 + \eta_2 - \eta_3 \right)
\label{eq:zeta}
\end{equation}
where
\begin{equation}
\eta_1 = \frac{\sigma_1}{\sigma_{\text{ref}}}, 
\qquad
\eta_2 = \frac{\sigma_2}{\sigma_{\text{ref}}} 
\qquad \text{and} \qquad
\eta_3 = \frac{\sigma_3}{\sigma_{\text{ref}}}.
\label{eq:etas}
\end{equation}

Two scenarios are possible within this scheme:
\begin{itemize}
\item \emph{Case~A} ($\sigma_1<\sigma_2$): 
\begin{equation}
\eta_1=1, \quad \eta_2=\frac{\sigma_2}{\sigma_1}\equiv \eta >1, \quad \eta_3=\frac{\sigma_3}{\sigma_1}, \quad \zeta = \frac{1}{2} \left(1 + \eta - \eta_3 \right), \quad \sigma_{\text{ref}}=\sigma_1.
\label{eq:etaSol1}
\end{equation}
\item \emph{Case~B} ($\sigma_1>\sigma_2$):
\begin{equation}
\eta_1=\frac{\sigma_1}{\sigma_2}\equiv \eta >1, \quad \eta_2=1, \quad \eta_3=\frac{\sigma_3}{\sigma_2}, 
\quad \zeta = \frac{1}{2} \left(1 + \eta - \eta_3 \right), \quad \sigma_{\text{ref}}=\sigma_2.
\label{eq:etaSol2}
\end{equation}
\end{itemize} 

In the following, we use variables $\eta >1$ and $0<\zeta<1$ to treat both cases simultaneously, since these two variables are sufficient to include all possible values of surface tensions.

\begin{figure}
\includegraphics[width=0.5\linewidth]{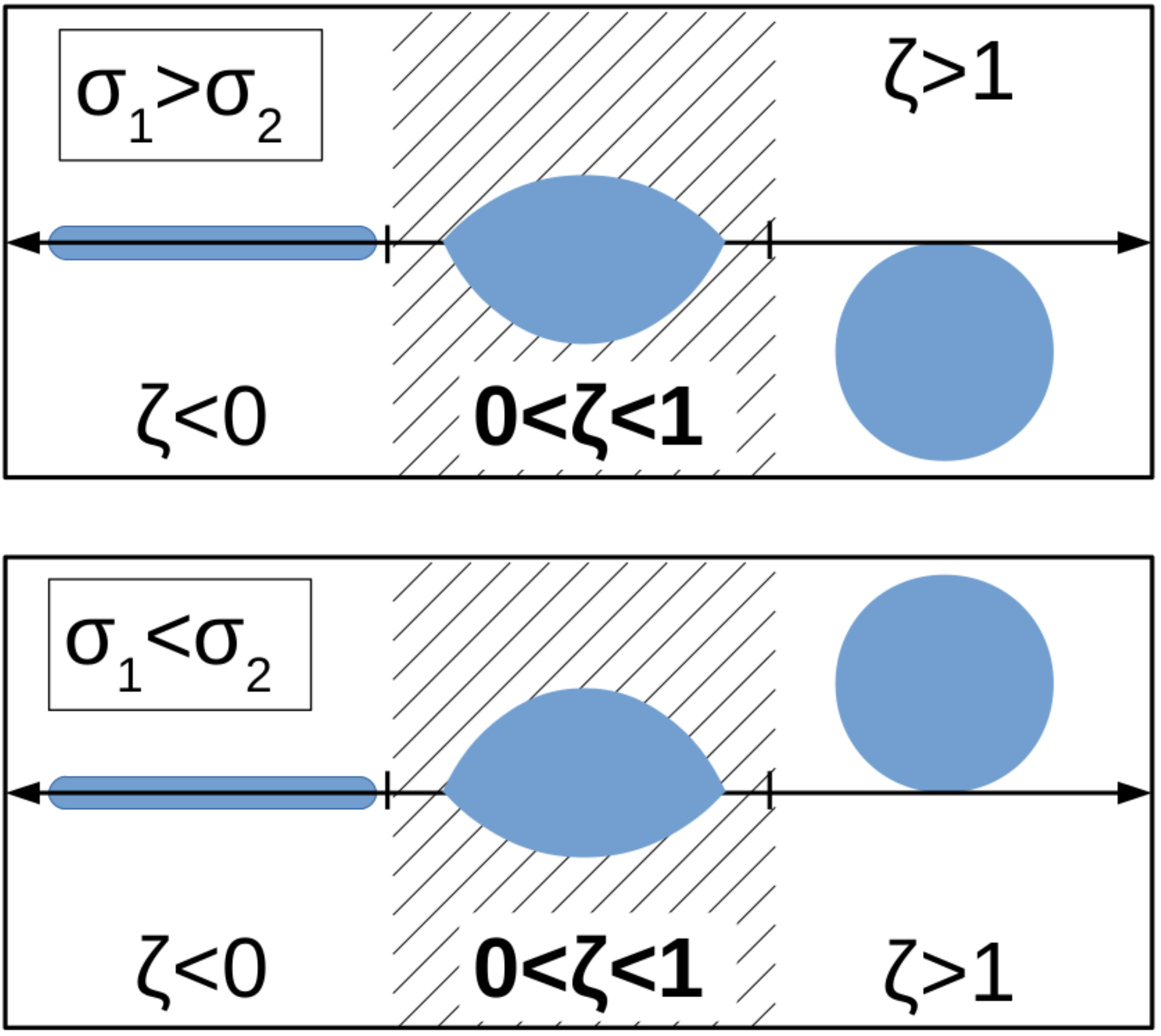}
\caption{Schematic of drop shapes for a given $\eta$ as function of $\zeta$, with $\sigma_1>\sigma_2$ (top panel) and $\sigma_1<\sigma_2$ (bottom panel). The central dashed area where $0<\zeta<1$ ($S^\ast<S<0$), corresponds to the partial wetting scenario studied in this work. For $\zeta<0$ ($S>0$), we have complete wetting, so that the drop spreads indefinitely. Instead, for $\zeta>1$ ($S<S^\ast$) a non--wetting case occcurs, where the drop remains on top or above the free surface, depending on the relative values of $\sigma_1$ and $\sigma_2$.}
\label{fig:spread_scheme}
\end{figure}

The different wetting possibilities given by Eq.~(\ref{eq:SCond}) are schematically shown in Fig.~\ref{fig:spread_scheme} in terms of the dimensionless parameters $\eta$ and $\zeta$. The left column in the figure represents $S>0$, where the drop spreads over the liquid surface. Therefore, complete wetting occurs for both case~A ($\sigma_1<\sigma_2$) and case~B ($\sigma_1>\sigma_2$). The right column represents  the non--wetting case, where the drop finds an equilibrium just below the interface for case~B and just above for case~A. Finally, in the center column, we have the partial wetting case, which is the scenario studied here.

By using these definitions, we can rewrite the three equilibrium equations, Eq.~(\ref{eq:equil_adim}), in dimensionless form as
\begin{subequations}
\begin{equation}
\Delta P_1 + \left(\frac{Bo}{\eta_1}\right) \left(\frac{\rho_A - \rho_C}{\rho_A}\right) z = \kappa_1
\end{equation}
\begin{equation}
\Delta P_2 + \left(\frac{Bo}{\eta_2}\right) \left(\frac{\rho_B - \rho_A}{\rho_A}\right) z = \kappa_2
\end{equation}
\begin{equation}
\Delta P_3 + \frac{Bo}{\eta_1 + \eta_2 - 2\zeta} \left(\frac{\rho_C - \rho_B}{\rho_A}\right) z = \kappa_3,
\end{equation}
\label{eq:sistAdim}
\end{subequations}
where 
\begin{equation}
Bo=\frac{\rho_A\, g\,  {\cal R}_0^2}{\sigma_{\text{ref}}}
\label{eq:BoReference}
\end{equation}
is the reference Bond number.

\section{Analytical solution without gravity}\label{sec:analyticalSol}

As a first attempt to solve this problem, we consider the case without gravity. One feature of this simplification is that the pressure $p_i$ is only determined by the reference pressure in each bulk, $P_i$, due to the absence of the bouyancy contribution. This fact implies that the reference pressures $P_B$ and $P_C$ must be equal at both sides of curve $3$ for all points along this curve, so that, $\kappa_3=0$ and curve $3$ is completely flat. With this condition $Bo = 0$, we have two simplified equations to be solved on the drop surfaces, 
\begin{equation}
\Delta P_i  = \kappa_i \quad \text{for} \quad i=1,2.
\label{Eq.NGDifEq}
\end{equation}

Since a flat curve $3$ implies $\gamma = 0$ (Fig.~\ref{fig:angles_scheme}), Eq.~(\ref{eq:NewmanRelations}) allows us to write $\alpha$ and $\beta$ in terms of the dimensionless parameters $\eta$ and $\zeta$ as 
\begin{subequations}
\begin{equation}
\alpha = \pi - \arccos \left[ \frac{\eta_2^2 - \eta_1^2 - (\eta_1 + \eta_2 - 2 \zeta)^2}{2 \eta_1 (\eta_1 + \eta_2 - 2 \zeta)} \right]\\
\end{equation}
\begin{equation}
\beta  = \arccos \left[1 + \frac{2\zeta (\zeta - \eta_1 - \eta_2)}{\eta_1 \eta_2} \right] - \alpha,
\end{equation}
\label{eq:AlfaBetaSG}
\end{subequations}
where $\eta_1$ and $\eta_2$ must be replaced by the corresponding values according to Eqs.~(\ref{eq:etaSol1}) and (\ref{eq:etaSol2}).

Moreover, Eq.~(\ref{Eq.NGDifEq}) shows that $\kappa_1$ and $\kappa_2$ are constants, so that the drop is formed by the intersection of two spherical caps whose radii of curvature are given by 
\begin{equation}
R_{1} = \frac{R_d}{\sin \alpha}, \qquad
R_{2} = \frac{R_d}{\sin \beta}.
\label{eq:R1R2}
\end{equation}
The dimensionless volume contribution of each spherical cap is obtained as function of $R_d$ and the corresponding angle $\alpha$ or $\beta$ as:
\begin{equation}
\begin{aligned}
V_{1} &= \frac{\pi}{6} R_d^3 \tan \frac{\alpha}{2} \left( 3 + \tan^3 \frac{\alpha}{2}\right) \\
V_{2} &= \frac{\pi}{6} R_d^3 \tan \frac{\beta}{2} \left( 3 +  \tan^3 \frac{\beta}{2}\right).
\end{aligned}
\label{eq:capsVol}
\end{equation}
Considering that $V_1 + V_2 = V = 4 \pi /3$ the drop radius is
\begin{equation}
R_d = 8^{1/3} \left[ 3 \left( \tan \frac{\alpha}{2} + \tan \frac{\beta}{2} \right) - \left( \tan^3 \frac{\alpha}{2} + \tan^3 \frac{\beta}{2} \right) \right]^{-1/3}.
\label{eq:RdSG}
\end{equation}

\begin{figure}[htb]
\subfigure[~Case A]{\includegraphics[width=0.45\linewidth]{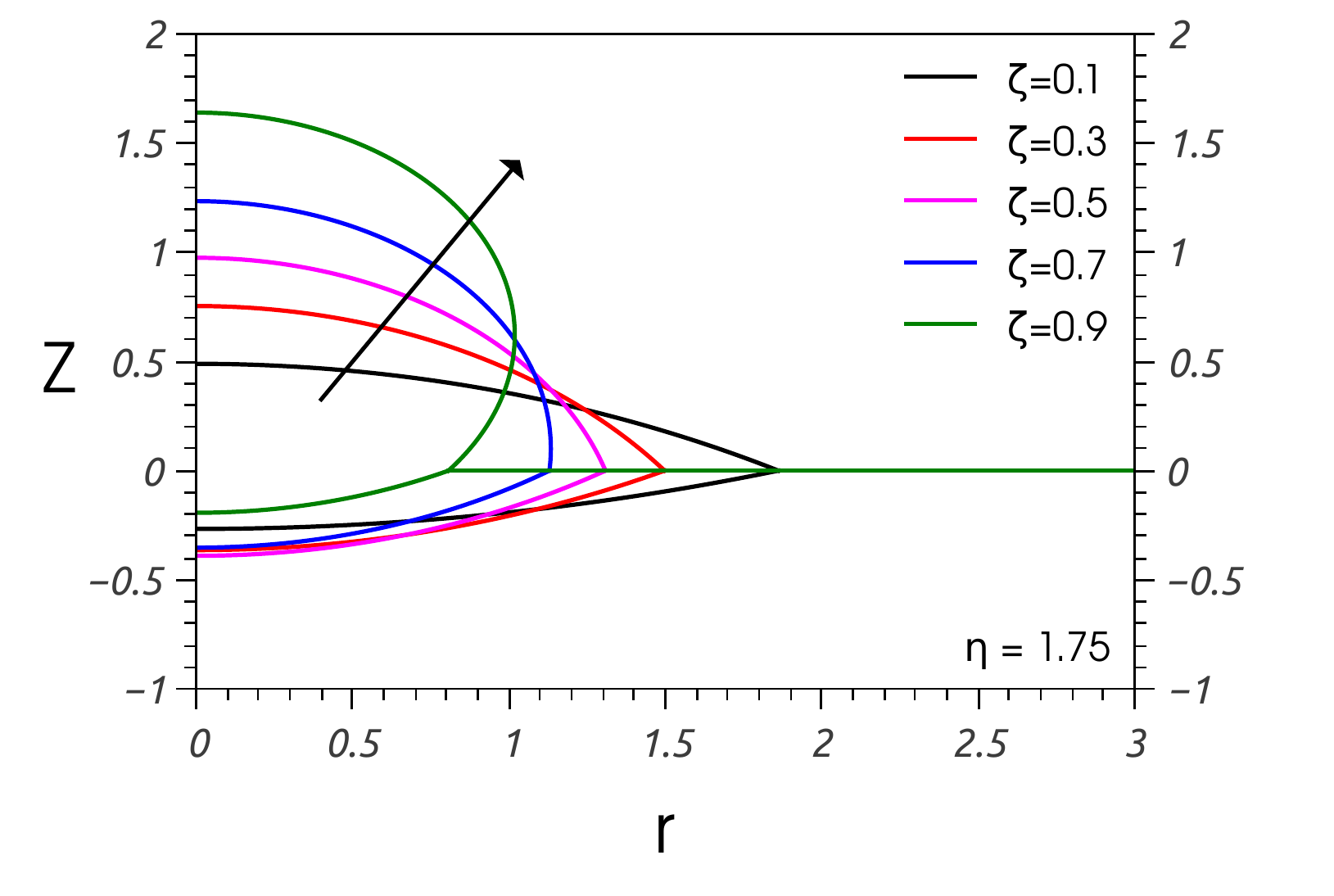}}
\subfigure[~Case B]{\includegraphics[width=0.45\linewidth]{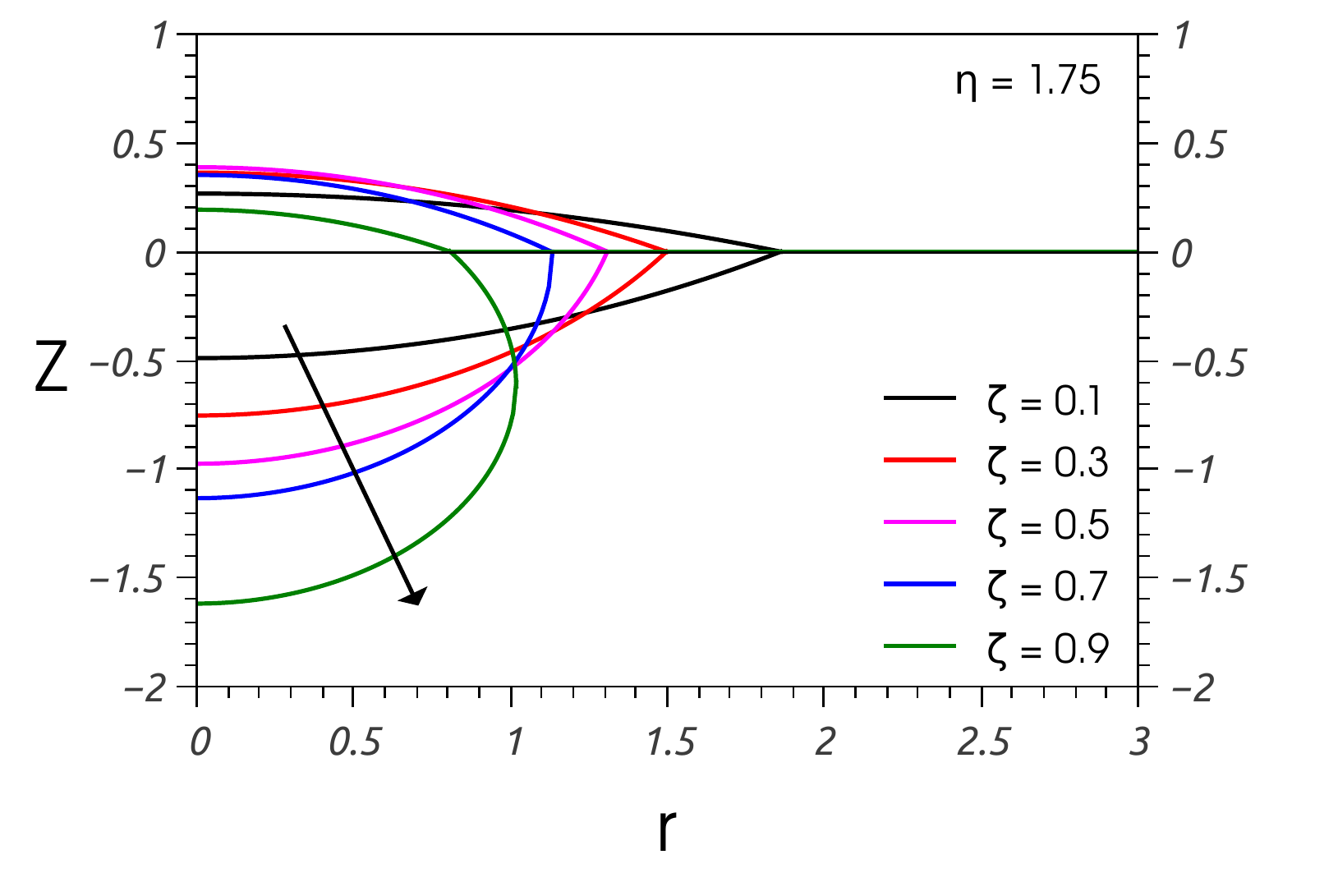}}
\caption{Solution without gravity: Drop profiles for $\eta = 1.75$ and $\zeta=0.1,0.3,0.5,0.9$ for cases A ($\sigma_1<\sigma_2$) and B ($\sigma_1>\sigma_2$). Note that the free surface of the liquid substrate (curve $3$) is flat for all $\zeta$.}
\label{fig:perfDifSolNG}
\end{figure}

In summary, for any value of $\sigma_{\text{ref}}$, Eqs.~(\ref{eq:AlfaBetaSG}), (\ref{eq:R1R2}) and (\ref{eq:RdSG}) allow calculation of the final equilibrium shapes for the entire range of possible values of $\eta$  and $\zeta$. For example, in Fig.~\ref{fig:perfDifSolNG} we show the drop profiles obtained for $\eta=1.75$ and $0.1 < \zeta < 1$ for case~A ($\sigma_1<\sigma_2$, Fig.~\ref{fig:perfDifSolNG}(a)) and case~B ($\sigma_2<\sigma_1$, Fig.~\ref{fig:perfDifSolNG}(b)). The drop height $h_d$ increases for both cases~A and B as $\zeta$ increases, mainly because of a $h_t$ (drop elevation) increase in case~A and a $h_b$ (drop sinking) increase in case~B, as was schematically presented in Fig.~\ref{fig:spread_scheme}. Interestingly, it can be shown that the lens shape of case~A (Fig.~\ref{fig:perfDifSolNG}(a)) corresponds to a reflection respect $z=0$ of case~B (Fig.~\ref{fig:perfDifSolNG}(b)), and vice versa. This is equivalent to changing the roles of Fluid~B and C (Fig.~\ref{fig:angles_scheme}), as can be seen in Eqs.~(\ref{eq:cos_law}) where, by exchanging $\sigma_1$ and $\sigma_2$, $\theta_A$ does not change while both $\theta_B$ and $\theta_C$ change sign. In spite of this result, we still show here the two cases, A and B, because the inclusion of gravity will break up this symmetry (see Sec.~\ref{sec:GravitySolutions}).

\begin{figure}[htb]
\subfigure[~Case A]{\includegraphics[width=0.45\linewidth]{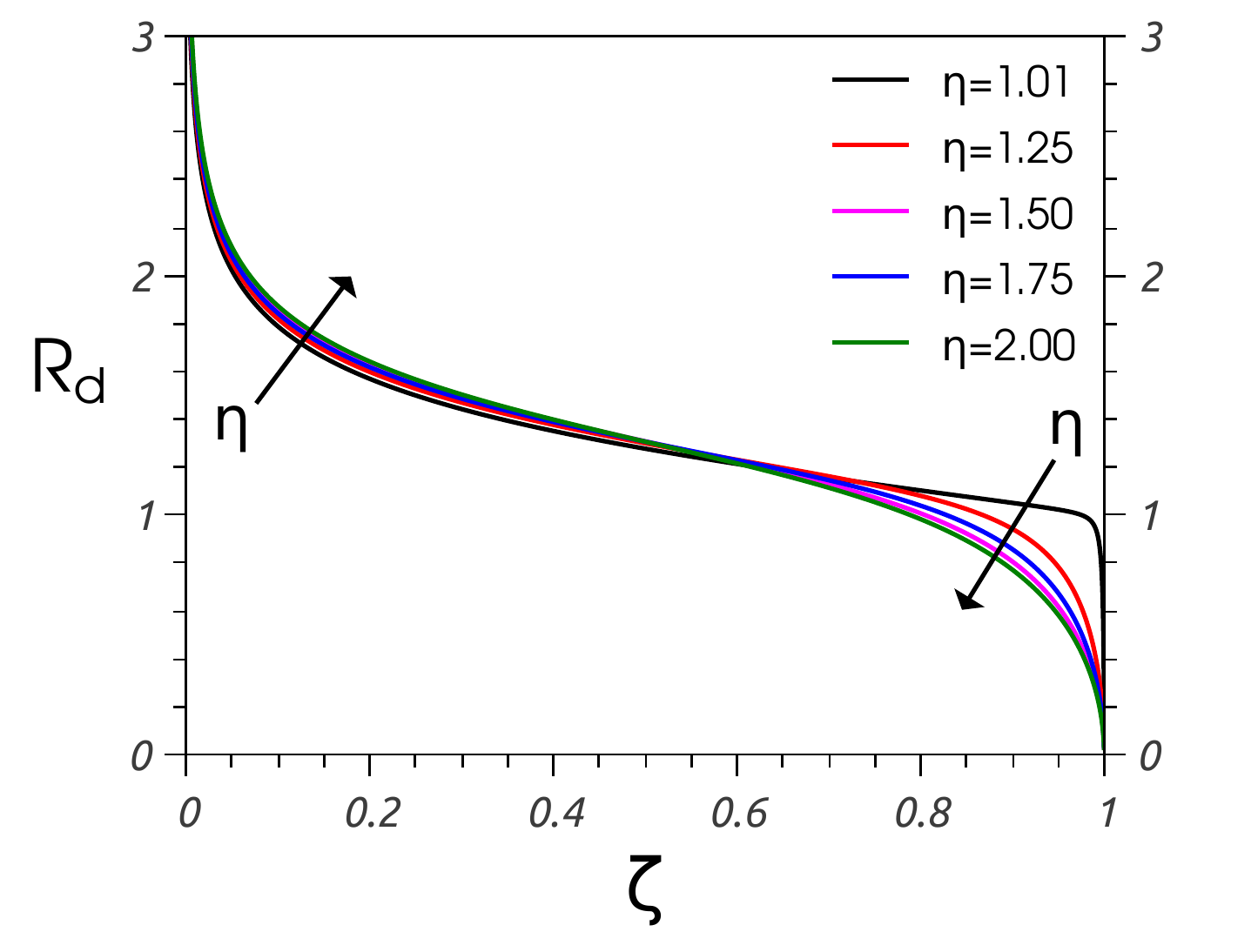}}
\subfigure[~Case B]{\includegraphics[width=0.45\linewidth]{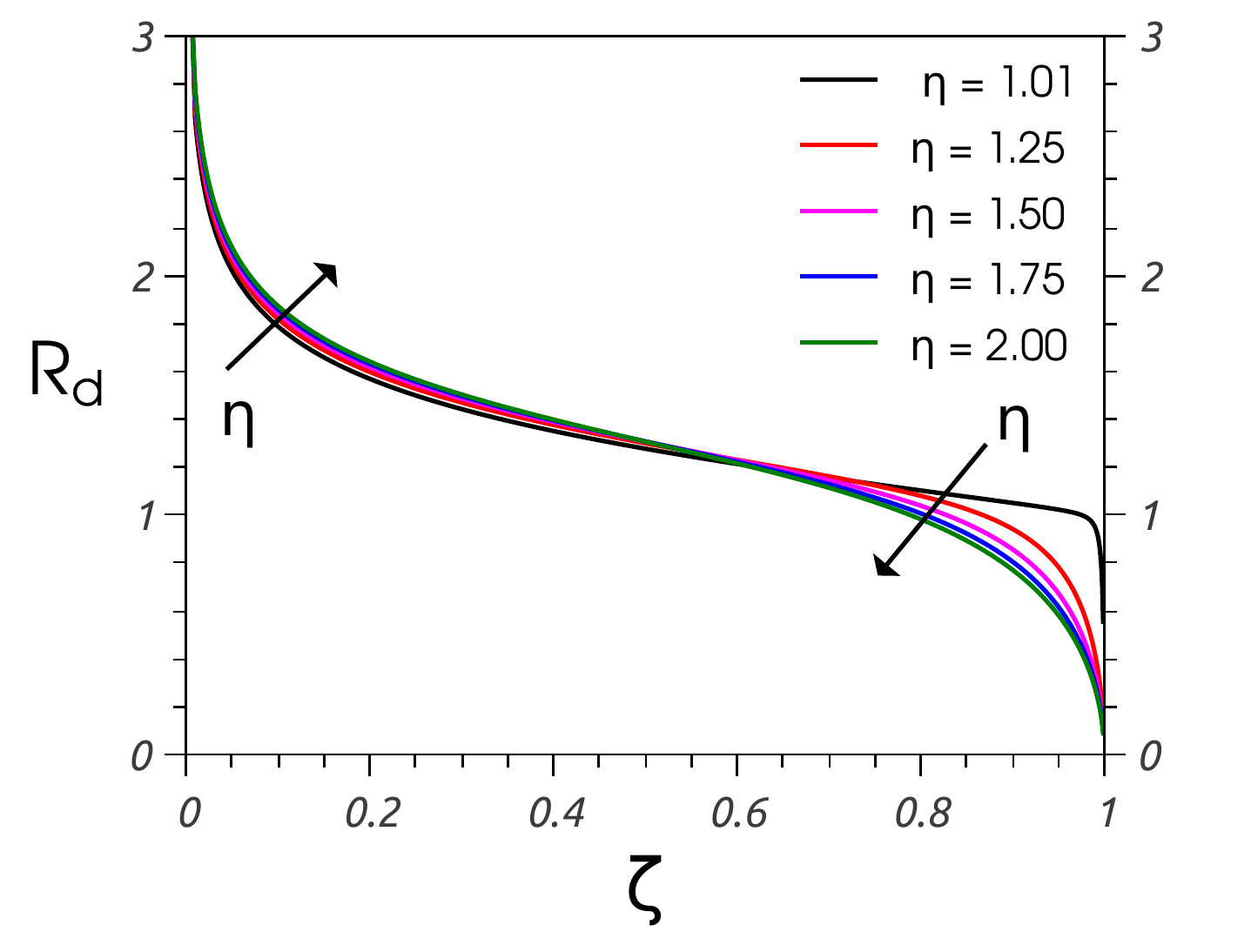}}
\caption{Solution without gravity: $R_d$ as function of $\zeta$ for $\eta = 1.01$, $1.25$, $1.50$, $1.75,2.00$ for cases A ($\sigma_1<\sigma_2$) and B ($\sigma_2<\sigma_1$). The arrows indicate the direction of increasing $\eta$.}
\label{fig:NGRdrop}
\end{figure}

To further characterize the solution without gravity, we show in Fig.~\ref{fig:NGRdrop} the drop radius, $R_d$, as a function of $\zeta$ for $\eta = 1.01$, $1.25$, $1.50$, $1.75$, $2.00$. We observe that for  both cases, $R_d$ decreases from large values at $\zeta \sim 0$ to $R_d =0$ at $\zeta = 1$ (recall that $\zeta=0$ corresponds to complete wetting), and that the $\eta$--dependence is only significant for $\zeta\gtrsim 0.8$. 

\begin{figure}[htb]
 \subfigure[~Case A]{\includegraphics[width=0.4\linewidth]{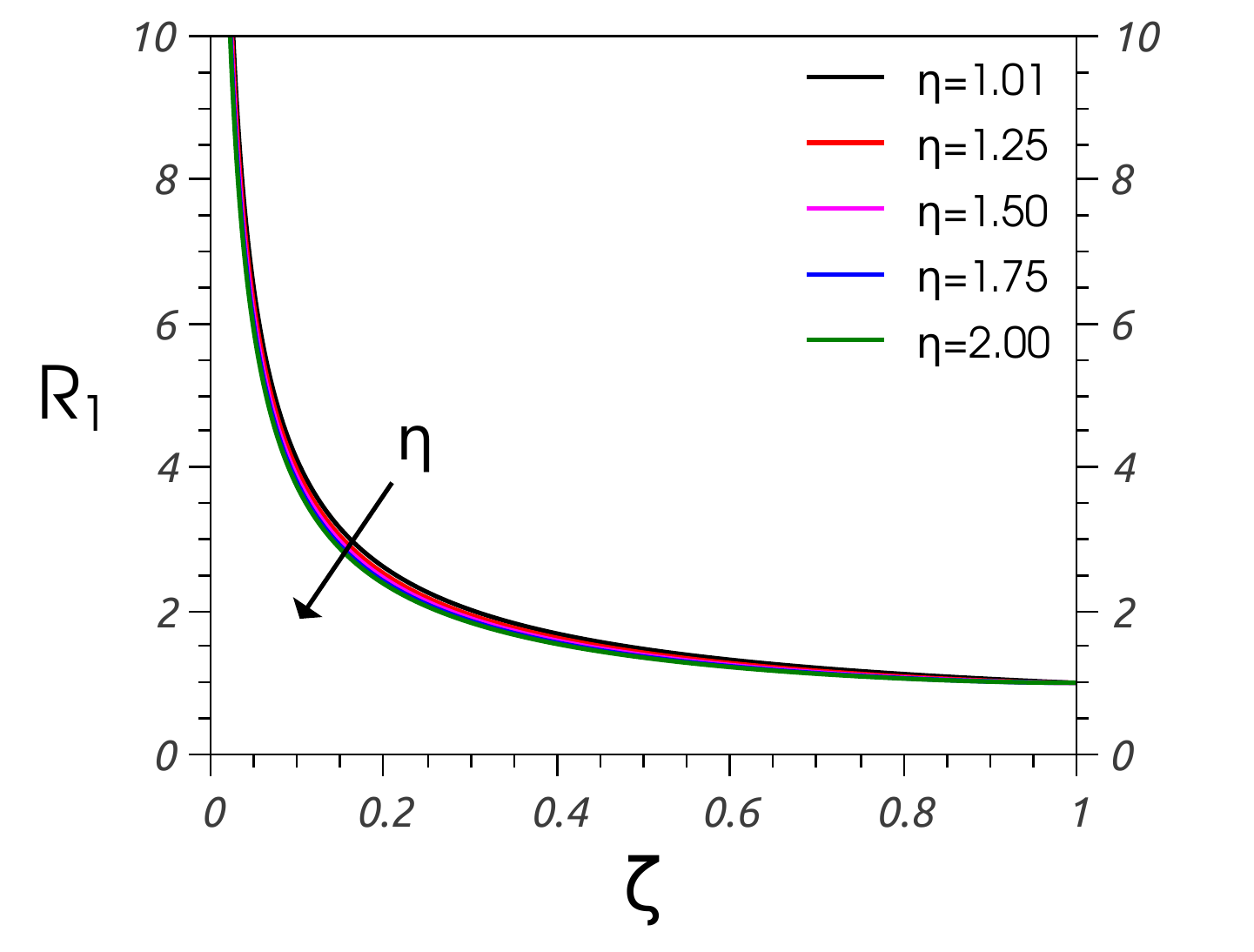}}
 \subfigure[~Case A]{\includegraphics[width=0.4\linewidth]{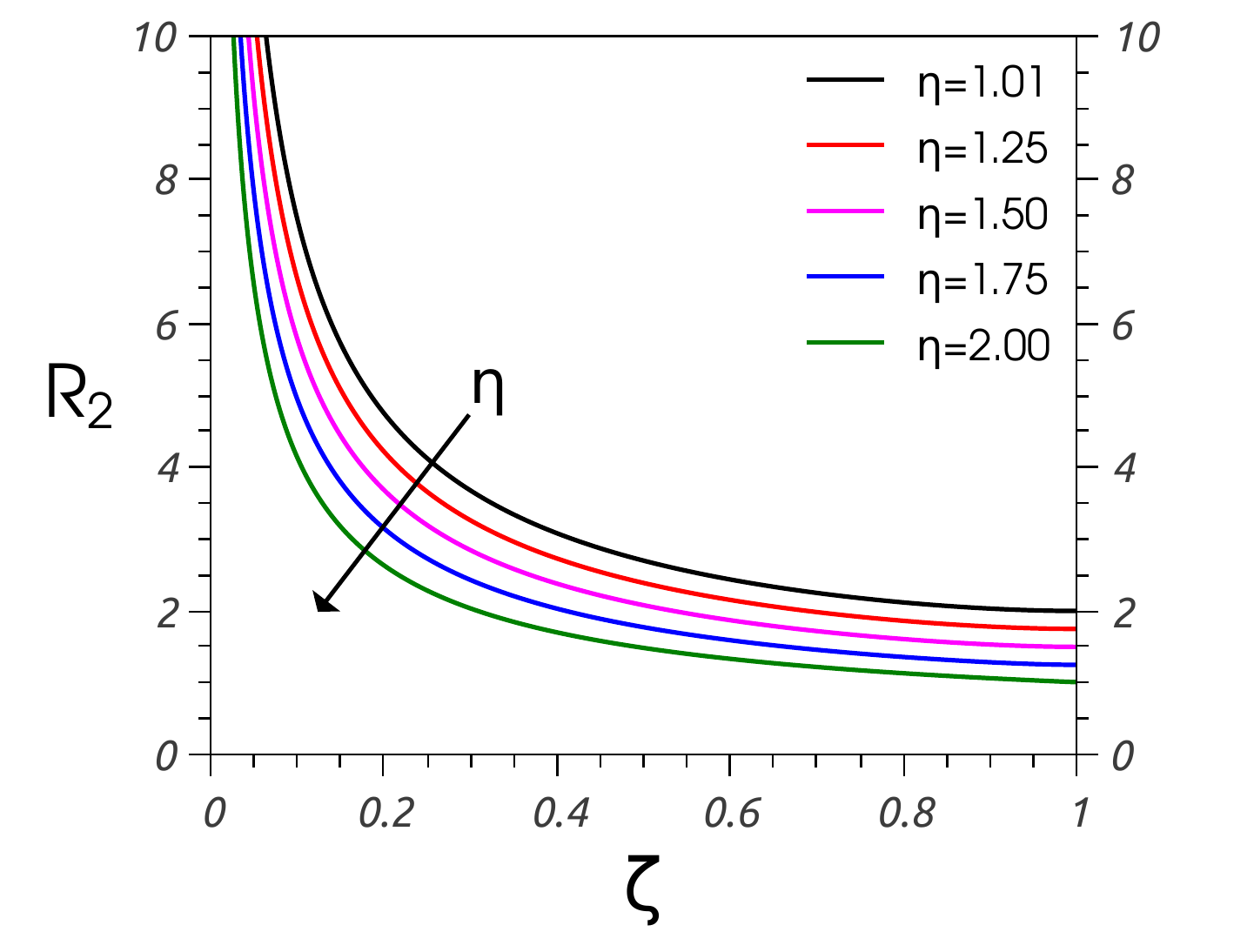}}\\
 \subfigure[~Case B]{\includegraphics[width=0.4\linewidth]{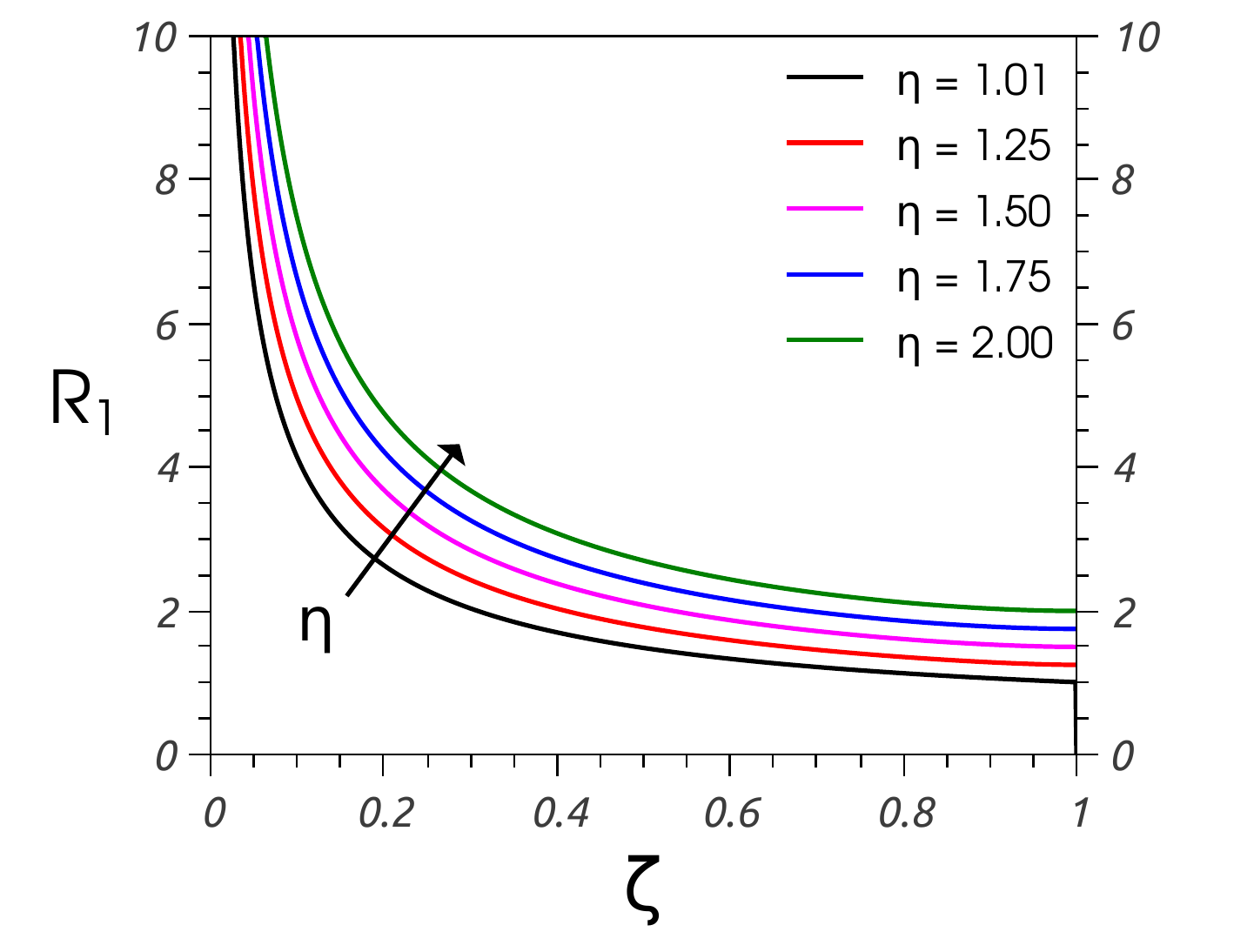}}
 \subfigure[~Case B]{\includegraphics[width=0.4\linewidth]{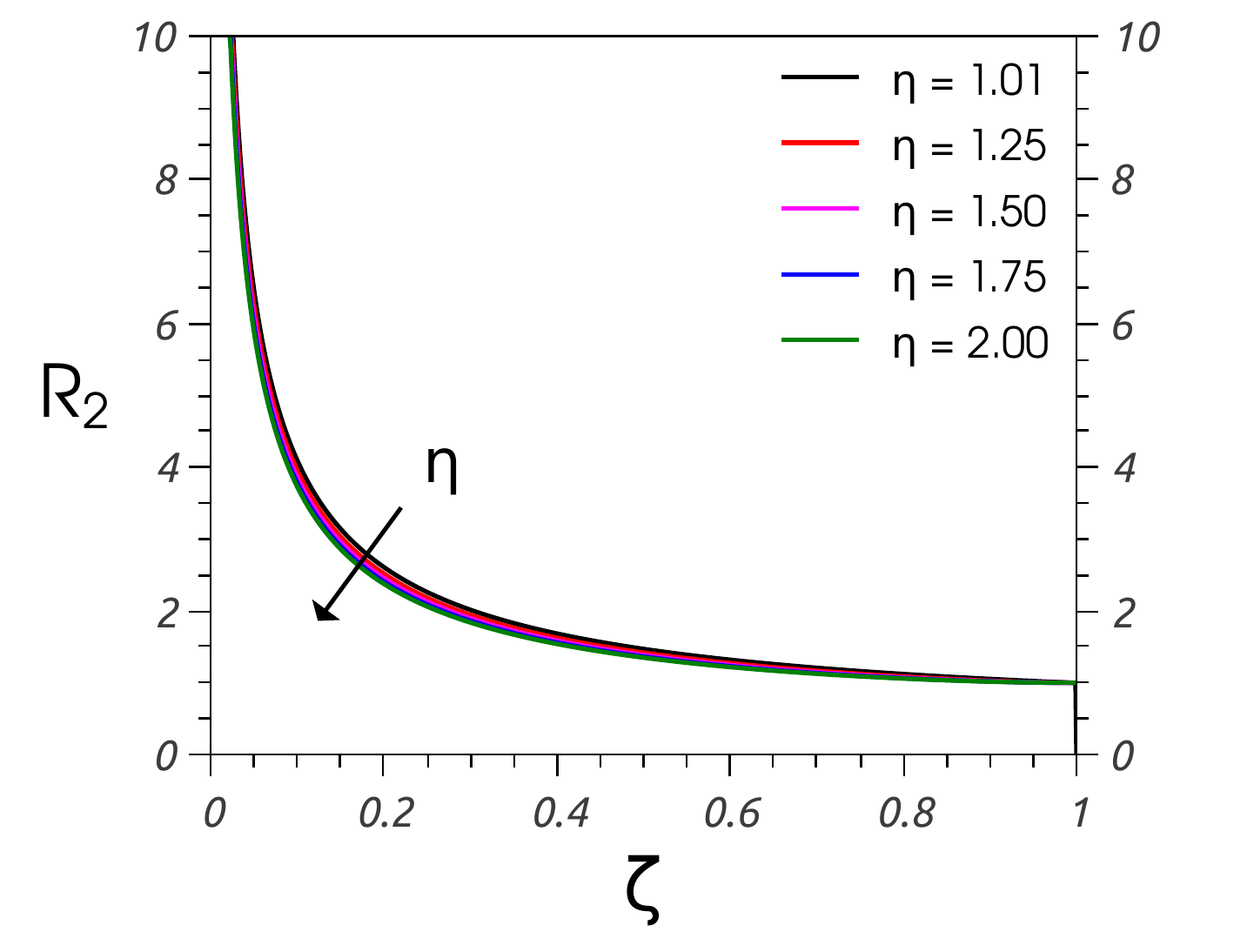}}
 \caption{Solution without gravity: Radius of curvature of each surface of the lens as a function of $\zeta$ for $\eta = 1.01$, $1.25$, $1.50$, $1.75$ and $2.00$: (a) For curve~1 and (b) curve~2 in case A ($\sigma_1<\sigma_2$) and (c) for curve~1 and (d) curve~2 in case B ($\sigma_2<\sigma_1$). The arrows indicate the direction of increasing $\eta$.}
 \label{fig:NGCurvRad}
 \end{figure}

The particular shape of the drop is defined by the radius of curvature of each spherical cap. In Fig.~\ref{fig:NGCurvRad} we show the radii $R_1$ of curve~$1$ and $R_2$ of curve~$2$ for $1<\eta<2$ and $0<\zeta<1$. In both cases these radii rapidly increase for $\zeta \rightarrow 0$ and approach a relatively flat zone as $\zeta \rightarrow 1$. We observe that $R_1$ in case~A and $R_2$ in case~B have the same behavior, while $R_2$ in case~A and $R_1$ in case~B also share an analogous behavior with the only difference that the dependence on $\eta$ occurs in opposite directions.

\section{Two possible solutions with gravity}\label{sec:GravitySolutions}

Unfortunately, it appears impossible to find an analytical solution of this three-phase problem with gravity. Therefore, we resort to the numerical solution of Eqs.~(\ref{eq:sistAdim}) with the corresponding dimensionless form of the conditions (Eqs.~(\ref{eq:rizi})--(\ref{eq:lens_vol})). To perform this task,  we develop an iterative scheme based on seven dimensionless variables, namely $(R_d, L_1, L_2, L_3, P_A, P_B, P_C)$ and we fix the length scale by choosing a volume ${\cal V}_0$ (see Eq.~(\ref{eq:R0})). In order to define the values of the reference Bond number, $Bo$, and the density factors in Eqs.~(\ref{eq:sistAdim}), we choose $\rho_A=0.97$~g$/$cm$^3$, $\rho_B=1.0$~g$/$cm$^3$, $\rho_C=0.0001$~g$/$cm$^3$ and ${{\cal V}_0}=0.02$~cm$^3$. With these values, we have $Bo = 1.22$ and $0.49$ for cases~A and B, respectively. We also choose $R_{wall}=2$~cm$/{\cal R}_0=11.88$.

To begin with, a first guess for $(L_1, L_2, L_3, R_d)$ can be taken from the analytical solution without gravity as obtained in Section~\ref{sec:analyticalSol}. Thus, $R_d^{(0)}$ is given by Eq.~(\ref{eq:RdSG}), and
\begin{equation}
L_1^{(0)} = R_1 \alpha, \qquad L_2^{(0)} = R_2 \beta, \qquad L_3^{(0)}=R_{wall}-R_d^{(0)},
\end{equation}
where $(\alpha, \beta)$ and $(R_1, R_2)$ are given by Eqs.~(\ref{eq:AlfaBetaSG}) and (\ref{eq:R1R2}), respectively.

\begin{figure}[htb]
\subfigure[~Case A]{\includegraphics[width=0.45\linewidth]{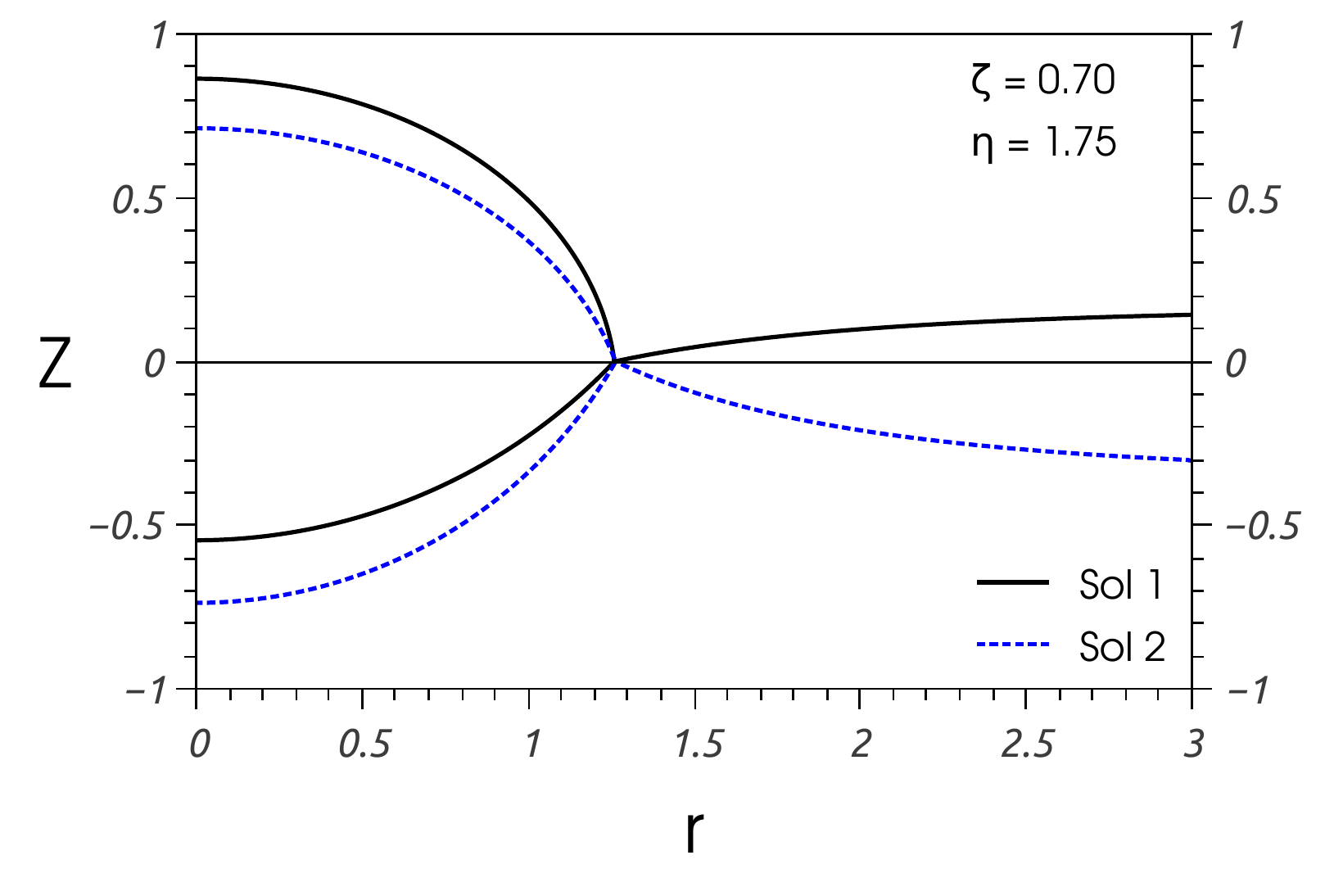}}
\subfigure[~Case B]{\includegraphics[width=0.45\linewidth]{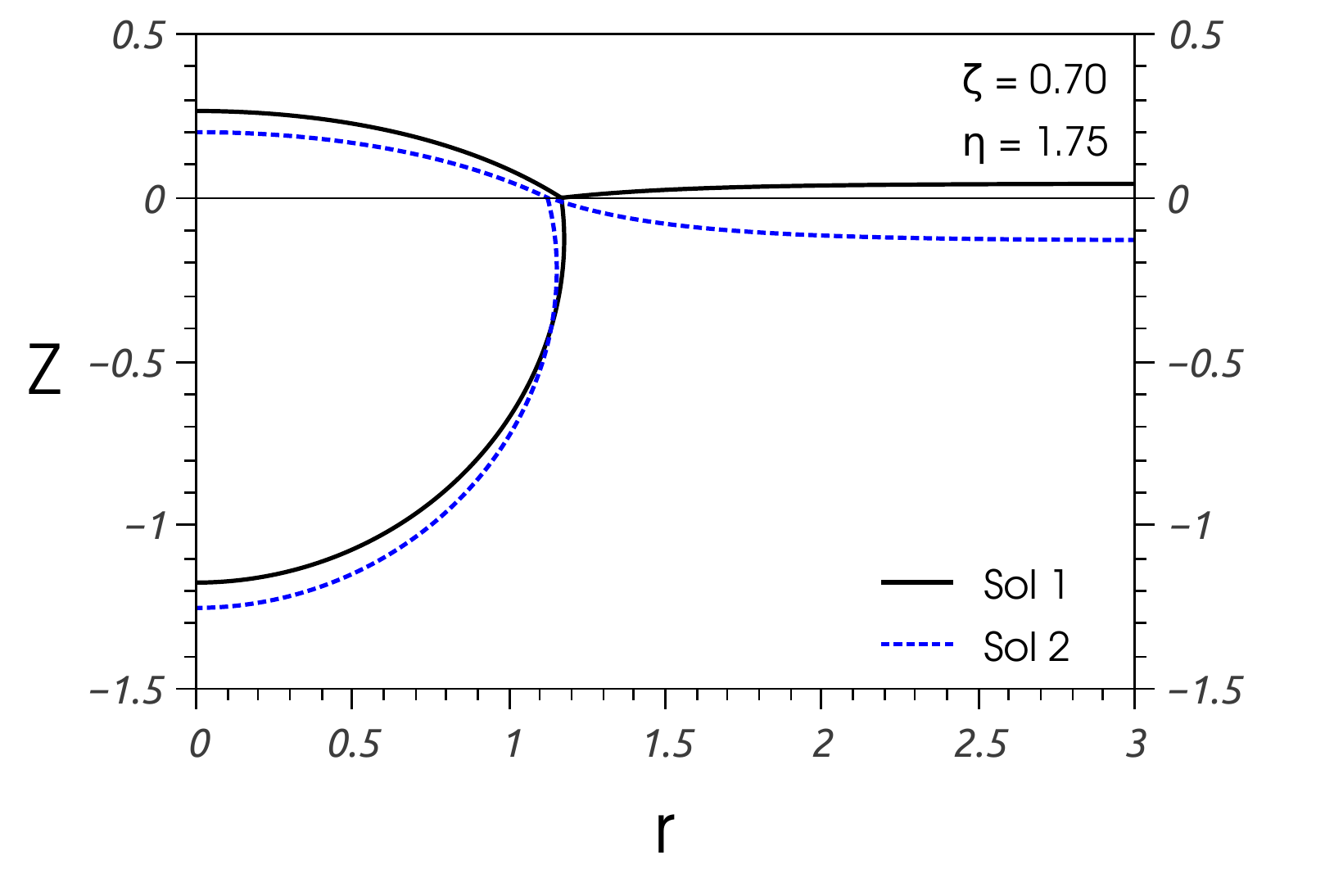}}
\caption{Example of the two families of solutions with gravity for $\eta=1.75$ and $\zeta = 0.70$. (a) Case~A ($\sigma_1 < \sigma_2$) and (b) case~B ($\sigma_2 < \sigma_1$).}
\label{fig:2solComp}
\end{figure}

Since the solution without gravity does not contain reference pressures, we have no available values to guess for for $(P_A, P_B, P_C)$. Here, we assume that $P_A^{(0)}$ and $P_B^{(0)}$ should be order one and of different signs because of the different orientation of the curvatures of curves~$1$ and $2$. Also, $P_C^{(0)}$ should be close to zero because we consider air as the surrounding fluid ($\rho_C \approx 0$). However, the signs of these variables cannot be guessed for given $\eta$ and $\zeta$ based on any plausible argument. We find that any choice of these variables can lead to one of these three possibilities namely, the solution does not converge or it may converge to two different types of solutions. An example of them, is shown in Fig.~\ref{fig:2solComp} for $\eta=1.75$, $\zeta=0.70$ and case~A (Fig.~\ref{fig:2solComp}(a)) or case~B (Fig.~\ref{fig:2solComp}(b)). In Appendix~\ref{ApendixGuess} we show that slight differences in the initial values of the parameters (mainly in the sign of $\Delta P_3$) could lead to any of these two possibilities. The existence of these two families, which we refer to as Sol $1$ and Sol $2$, do not depend on $\sigma_{\text{ref}} = \min (\sigma_1, \sigma_2)$ choice, since we find that these two families exist for $\sigma_1 < \sigma_2$ as well as for $\sigma_1 > \sigma_2$.

\begin{figure}[htb]
\subfigure[~Sol. 1]{\includegraphics[width=0.45\linewidth]{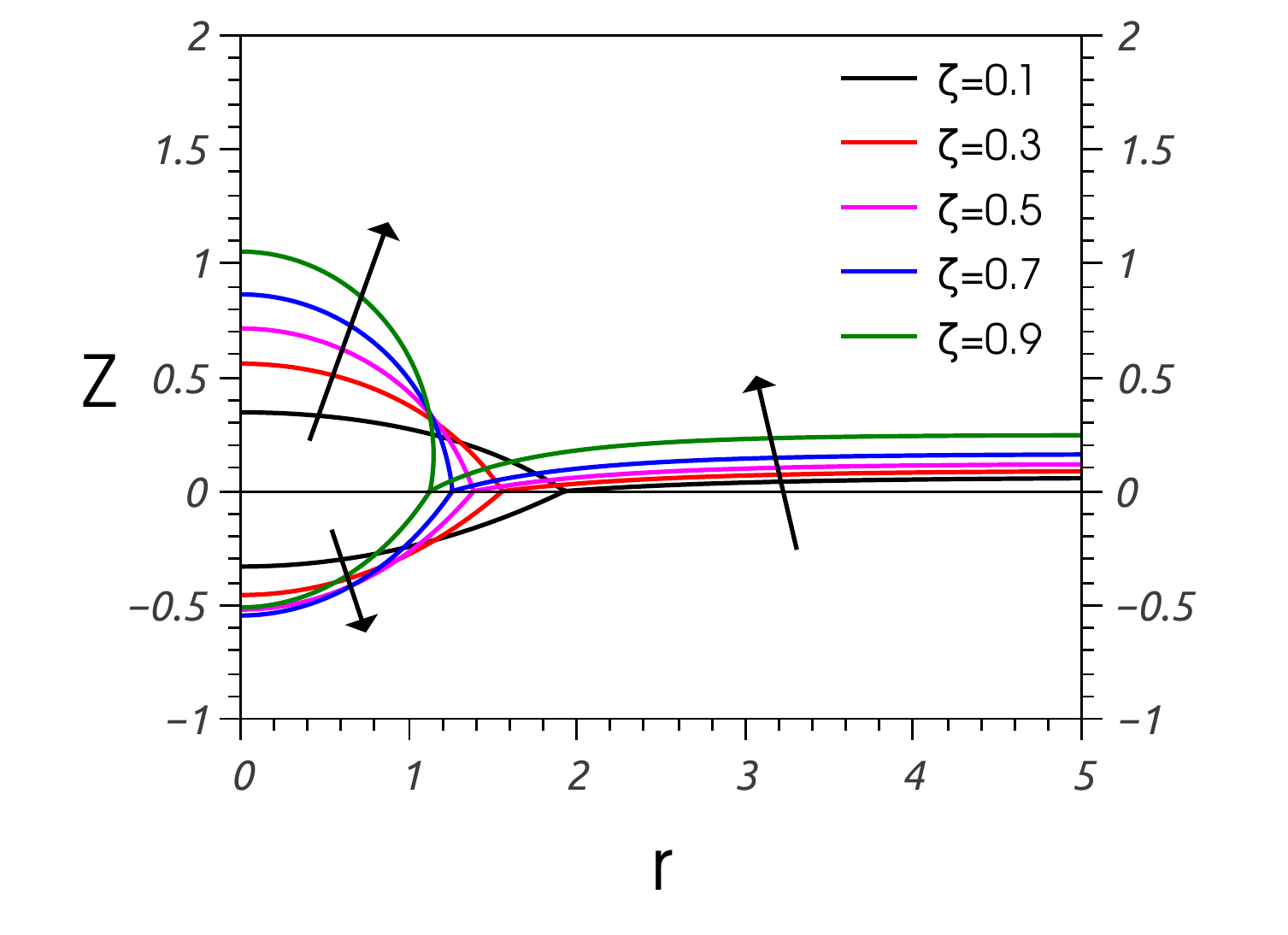}}
\subfigure[~Sol. 2]{\includegraphics[width=0.45\linewidth]{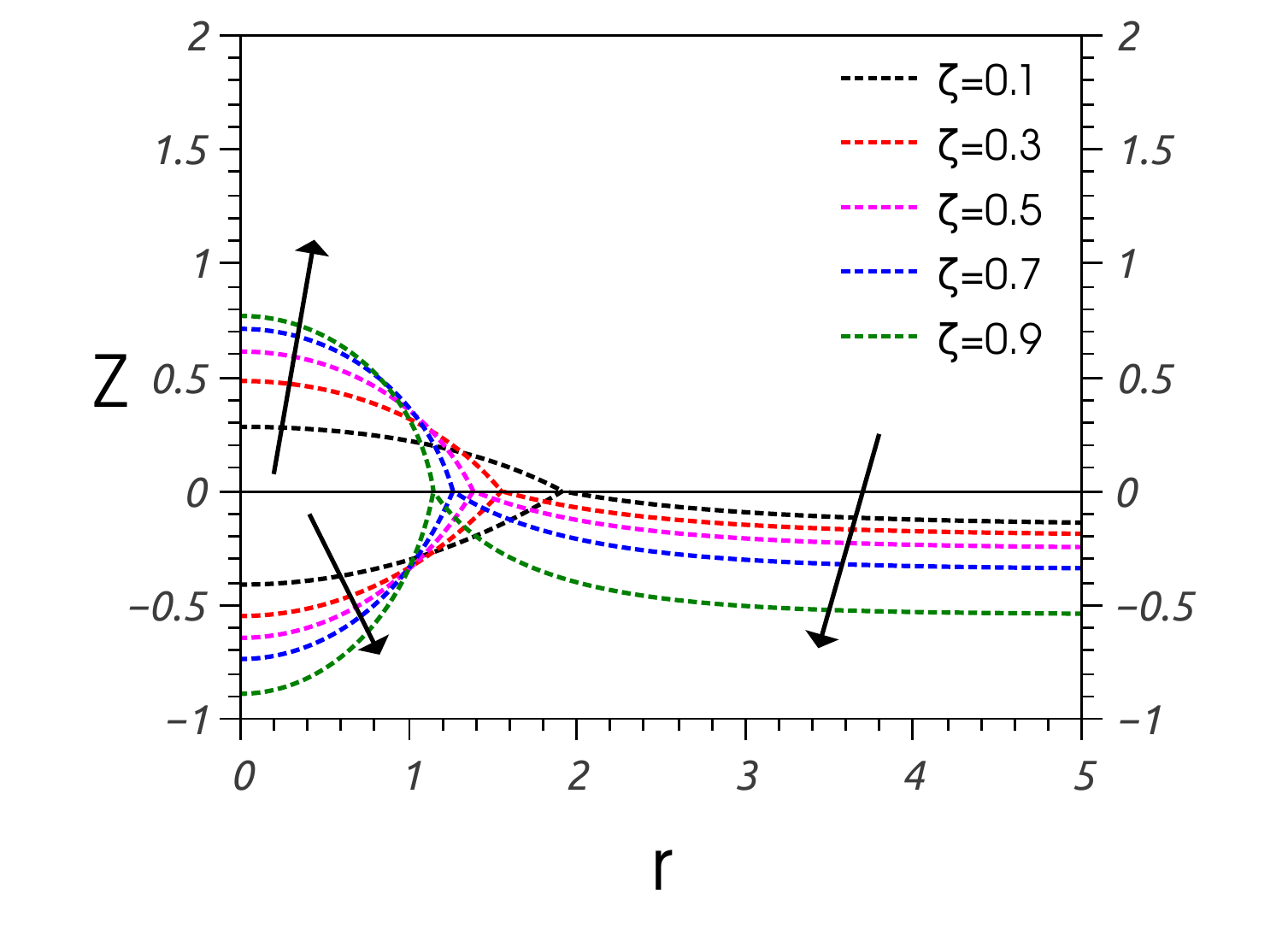}}
\caption{Case~A: Two families of solutions with gravity for $\eta=1.75$ and several values of $\zeta$. (a) Solutions type~$1$. (b) Solutions type~$2$. The arrows indicate the direction of increasing $\zeta$.}
\label{fig:perfDifSol}
\end{figure}

For case~A, the definitions in Eq.~(\ref{eq:etaSol1}) hold. In Fig.~\ref{fig:perfDifSol} we show examples of the two families of solutions for  $\eta=1.75$ and several $\zeta$'s. The first family (Sol~1 in Fig.~\ref{fig:perfDifSol}(a)) shows that the triple contact point is always under curve $3$, while in the second family (Sol~2 in Fig.~\ref{fig:perfDifSol}(b)) it is always above. Also, the shape of curve $3$ near the triple contact point has a different sing of the curvature for each type of solution. In both cases, the drop radius decreases as $\zeta$ increases. It is interesting to note that as $\zeta$ increases curve $3$ increases its curvature. 

\begin{figure}[htb]
\subfigure[~Sol. 1 ($Bo=0.49$)]{\includegraphics[width=0.45\linewidth]{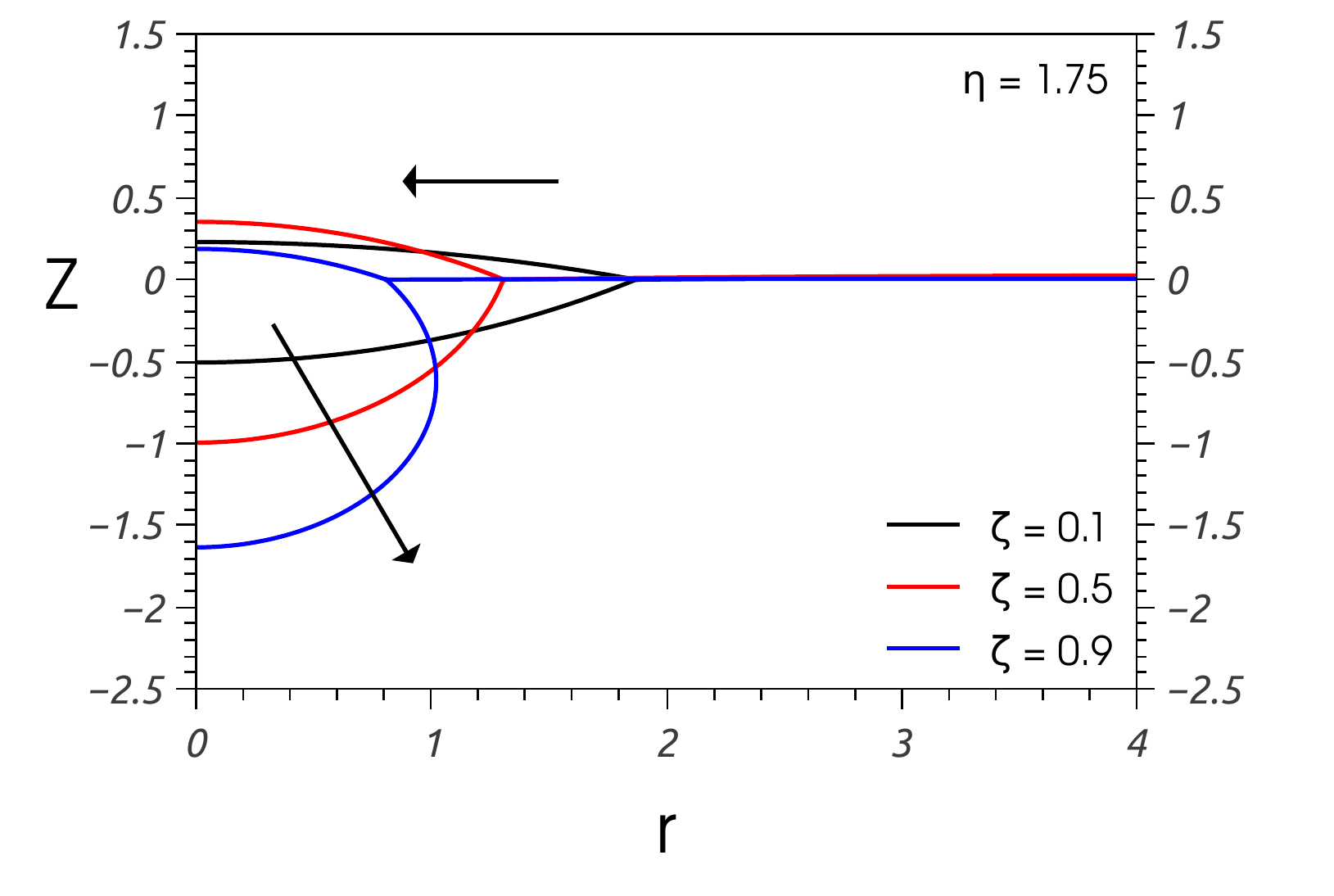}}
\subfigure[~Sol. 2 ($Bo=0.49$)]{\includegraphics[width=0.45\linewidth]{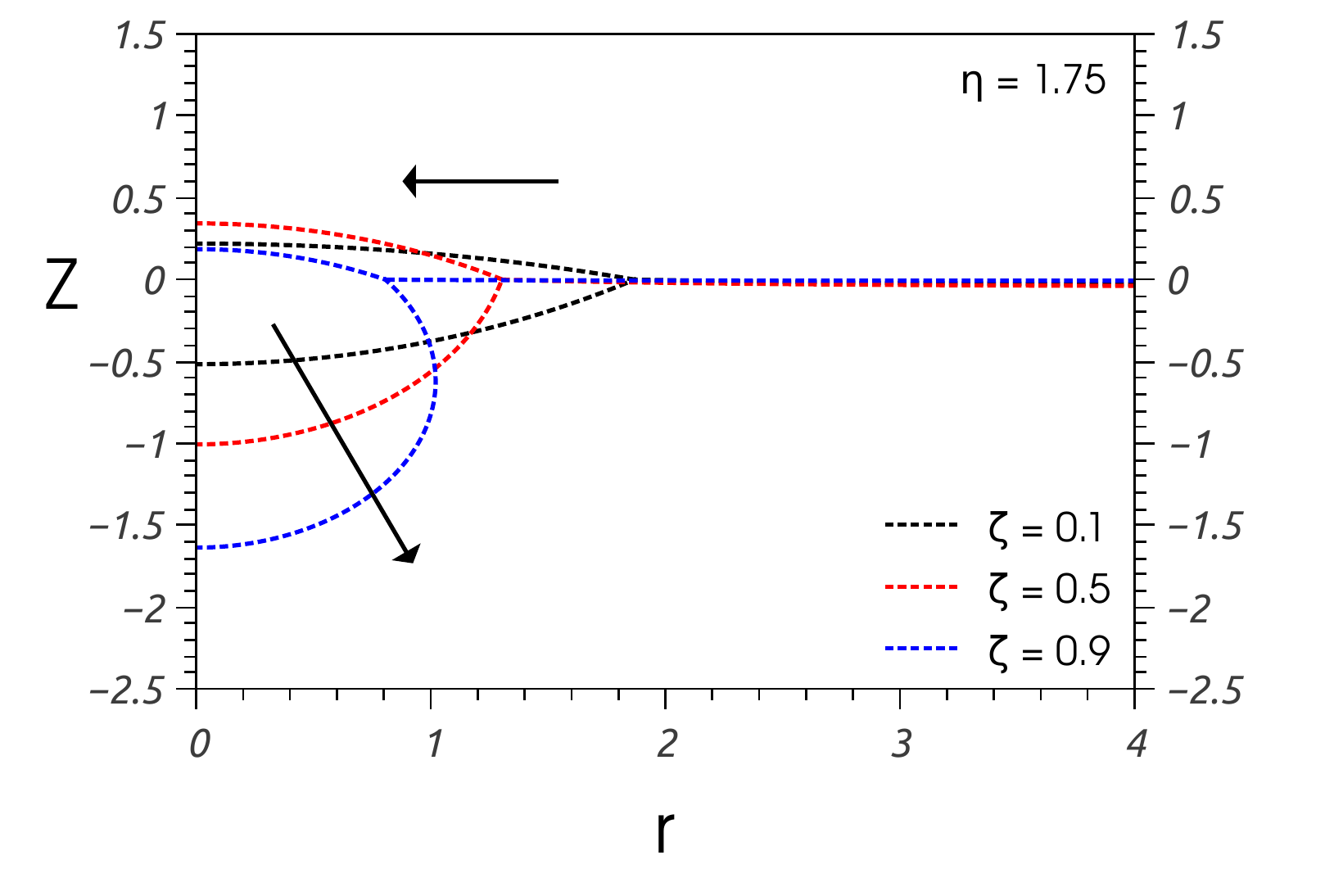}}\\
\subfigure[~Sol. 1 ($Bo=4.32$)]{\includegraphics[width=0.45\linewidth]{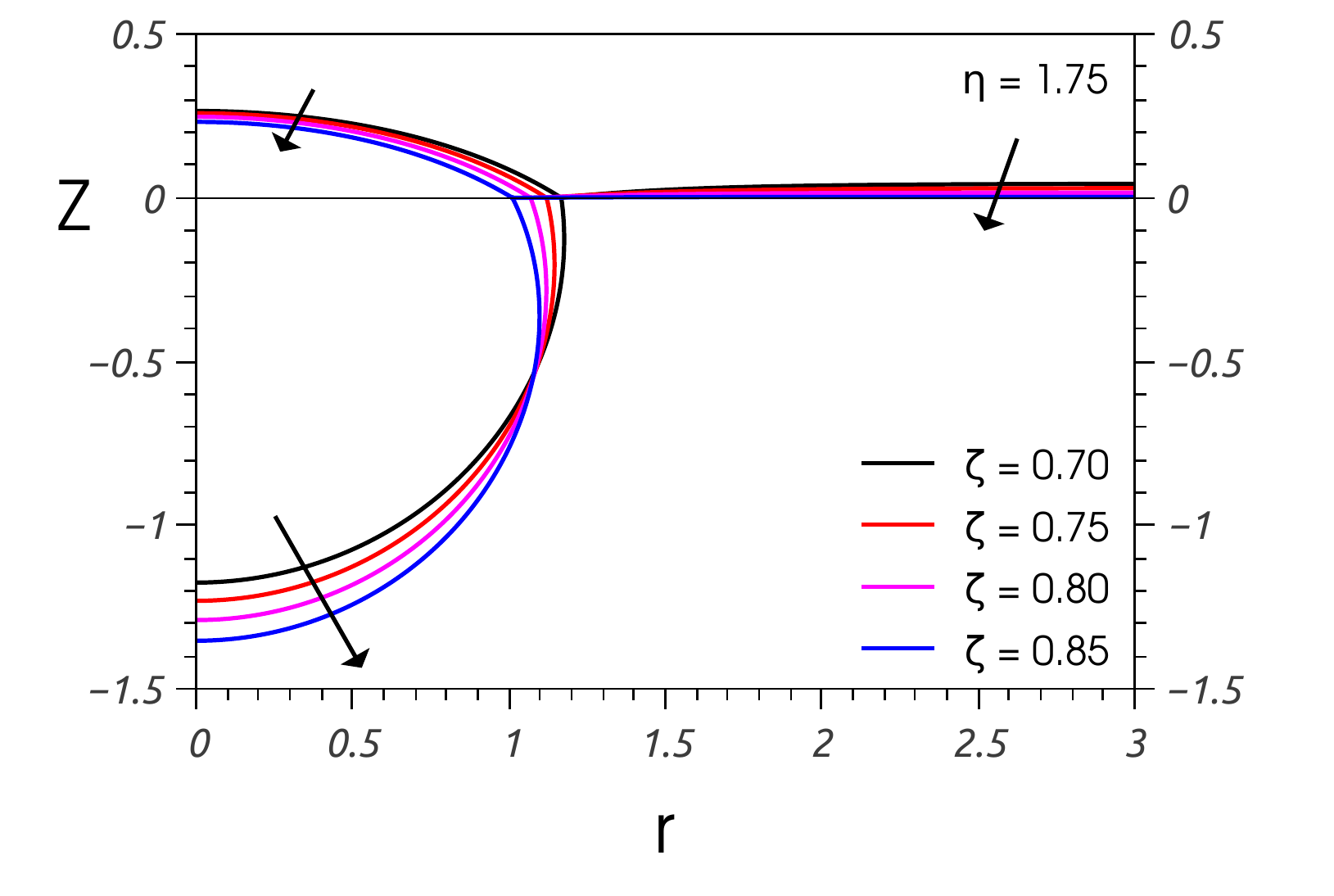}}
\subfigure[~Sol. 2 ($Bo=4.32$)]{\includegraphics[width=0.45\linewidth]{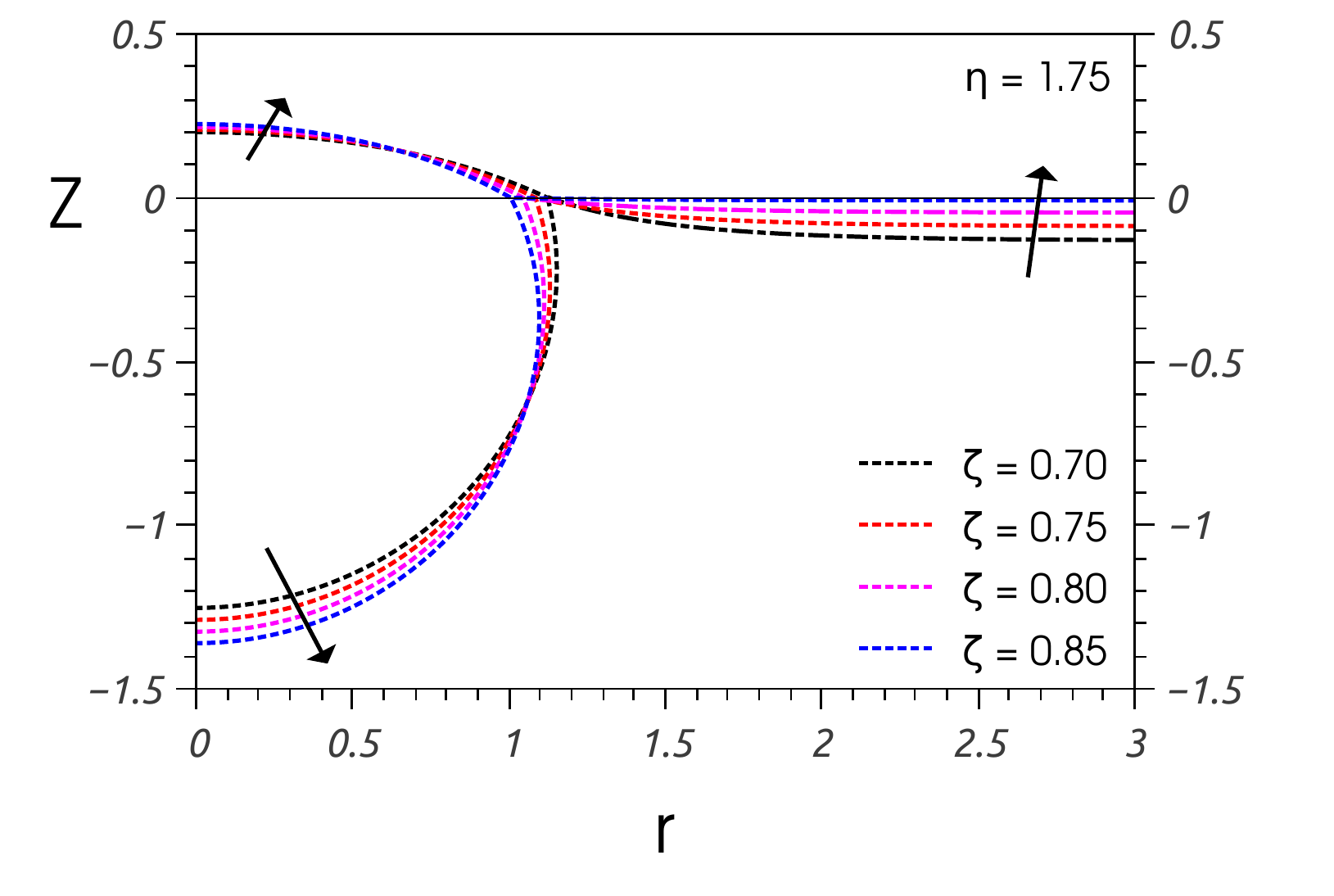}}
\caption{Case~B: Two families of solutions with gravity for $\eta=1.75$ and several values of $\zeta$. (a) Solutions type~$1$ ($Bo=0.49$). (b) Solutions type~$2$ ($Bo=0.49$). (c) Solutions type~$1$ ($Bo=4.32$). (d) Solutions type~$2$ ($Bo=4.32$). The arrows indicate the direction of increasing $\zeta$.}
\label{fig:perfDifSolsigmaRef2}
\end{figure}

Case~B requires the use of the definitions in Eq.~(\ref{eq:etaSol2}). In Fig.~\ref{fig:perfDifSolsigmaRef2} we show examples of the two possible solution families for $\eta=1.75$ and several values of $\zeta$. As before, the drop radius decreases as $\zeta$ increases in both cases. However, there are few differences: in this case it is possible to have $\beta$ larger than $\pi /2$, and both families converge to only one solution as $\zeta$ increases. Since the differences between Sol 1 and 2 are not clearly visible in Figs.~\ref{fig:perfDifSolsigmaRef2}(a) and (b), we show them also for a larger volume in Figs.~\ref{fig:perfDifSolsigmaRef2}(c) and (d), where ${{\cal V}_0} = 0.5$~cm$^3$ ($Bo=4.32$). 

\begin{figure}[htb]
\subfigure[]{\includegraphics[width=0.45\linewidth]{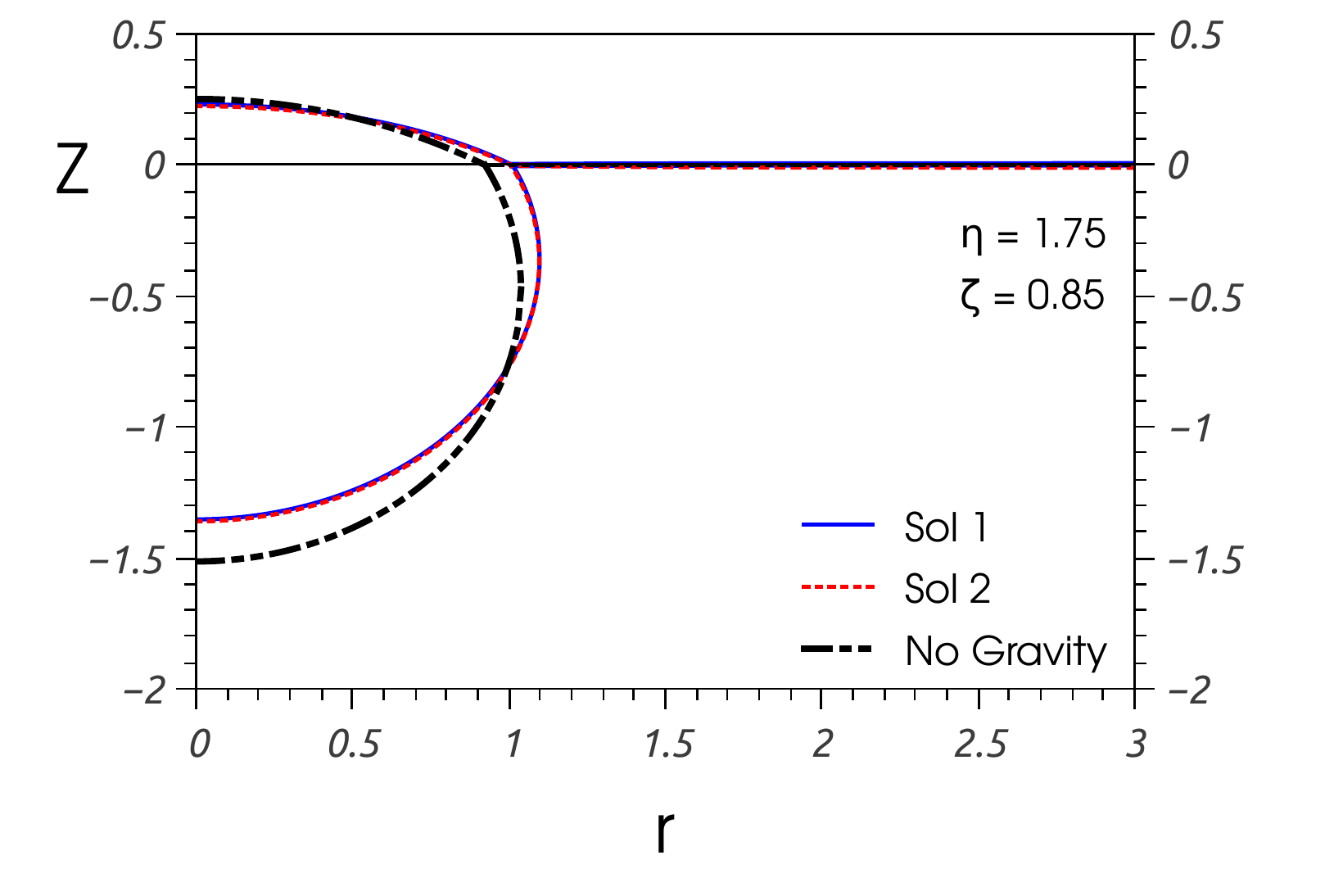}\label{fig:perfsigmaRef2NG}}
\subfigure[]{\includegraphics[width=0.45\linewidth]{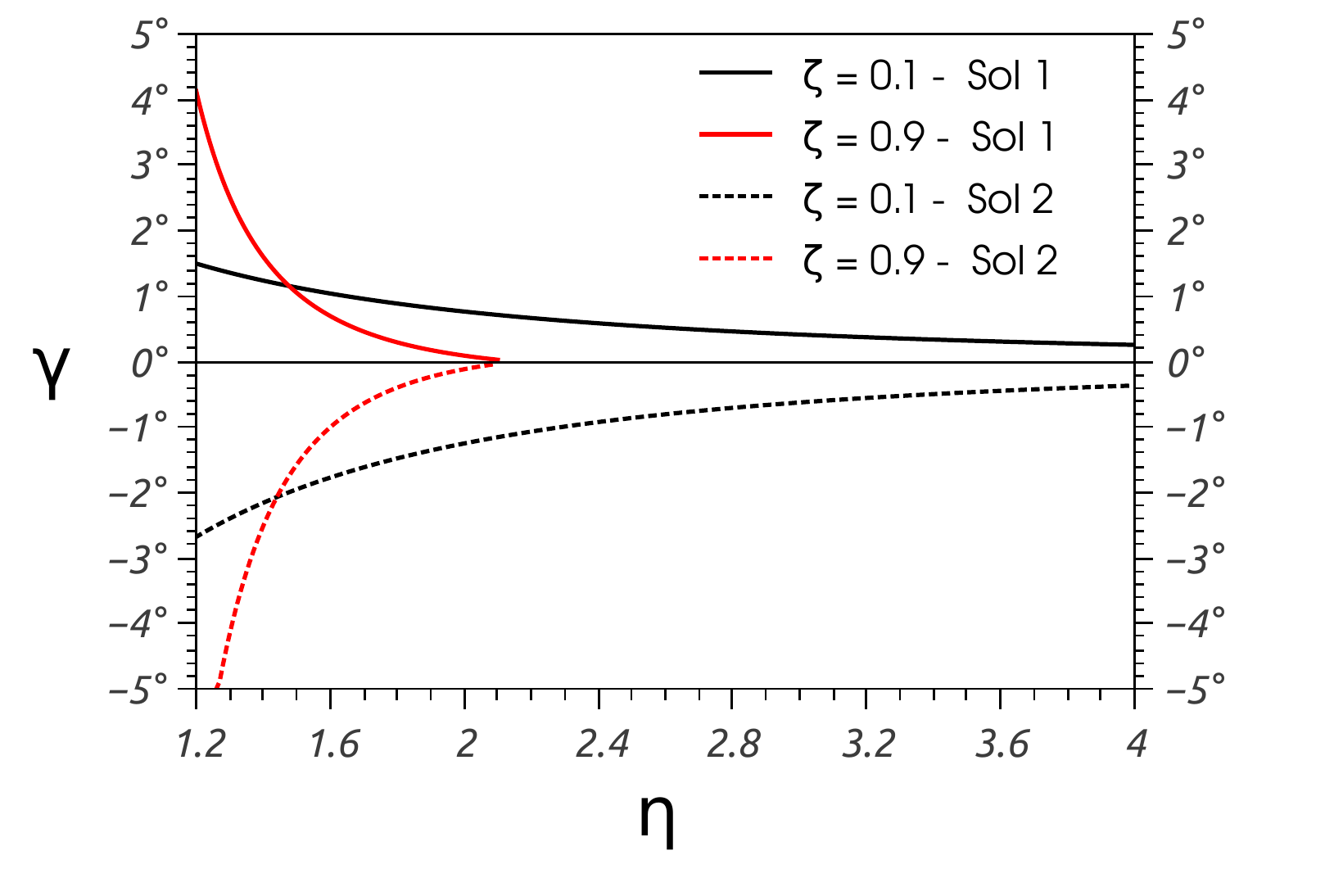}\label{fig:gamma}}
\caption{Case~B: (a) Two families of solutions with gravity for $\eta=1.75$, $\zeta=0.85$ and $Bo=4.32$ (practically coincident) compared with the analytical solution without gravity. (b) Evolution of angle $\gamma$ as function of $\eta$ for both Sol $1$ and Sol $2$ for two extreme values of $\zeta$ and $Bo=0.49$. As $\eta$ increases $\gamma$ decreases toward zero. The convergence to a flat Curve $3$ is faster for larger $\zeta$.}
\end{figure}

Moreover, we observe a limiting solution for both Sol~$1$ and $2$ at a certain pair of values $(\zeta, \eta)$ for which the curve~$3$ is practically flat, where both solutions become coincident. For the specific case in Figs.~\ref{fig:perfDifSolsigmaRef2}(c) and (d), these coincident solutions are shown in Fig.~\ref{fig:perfsigmaRef2NG}, where they are practically indistinguishable one from another (the solution without gravity for the same $\eta$ and $\zeta$ is also shown for comparison). In this case, the corresponding pair is $\zeta \approx 0.85$ and $\eta=1.75$. Note that the three solutions have a flat curve $3$ and $\beta> \pi/2$,  while the no--gravity solution has smaller $R_d$ and larger $h_d$ than the gravitational solutions.
 
Note also that this convergence of solutions towards a single solution with flat curve $3$ corresponds to $\gamma \rightarrow 0$. The behaviour of this angle can be used as a probe to observe the differences between Sol~$1$ and Sol~$2$. In fact, in Fig.~\ref{fig:gamma} we plot $\gamma$ as a function of $\eta$ for two extreme values of $\zeta$ as obtained for Solutions $1$ and $2$ for $Bo=0.49$. Clearly, their convergence occurs faster for larger $\zeta$ and smaller $\eta$. 

Unfortunately, we do not have mathematical or physical arguments to explain why there are two different converged solutions for the same physical parameters. Clearly, this topic deserves further study, but it is beyond the scope of the present work. Therefore, we proceed by calculating the total energy of each family under the expectation that the one with lower energy is actually the one to be observed in nature.

\section{Energy analysis}\label{sec:energy}

In order to analyze which of the equilibrium solutions identified in the previous section has lower energy, we calculate the total energy of each family of solutions for the problem with gravity as a function of $\eta$ and $\zeta$ for a fixed drop volume of ${\cal V}=0.2$~cm$^3$ and the corresponding $Bo$. We consider a region of finite size (i.e., a vessel, like in an experimental situation) in order to have finite values of the  total energy, where we include both surface and volumetric (gravity) contributions. Then, we calculate the difference between the total energy of the solution, $E$, and that of the system without the drop in it, $E_0$.

\begin{figure}[htb]
\includegraphics[width=0.6\linewidth]{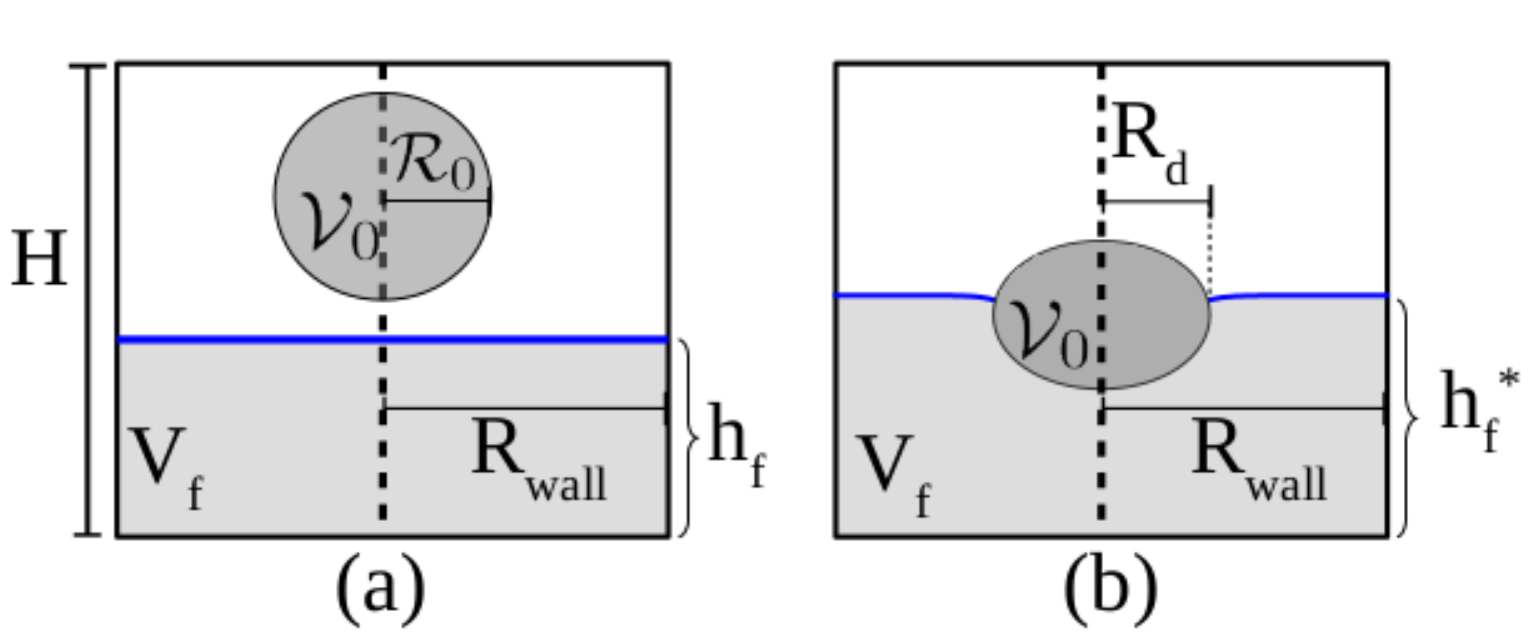}
\caption{(a) Initial and (b) final states of the approach used to calculate the system energy: it corresponds to a finite volume configuration with fixed $V_f$ and variable $h_f$.}
\label{fig:esqCalcEnerg}
\end{figure}

The integration domain is depicted in Fig.~\ref{fig:esqCalcEnerg}, which is a cylindrical container of radius $R_{wall}$ and height $H$. The level of fluid $B$ changes from $h_f$ to $h_f^\ast$ as the drop is deposited on its surface, since no flow is allowed through the vessel walls ($V_{B}=$const.).
The total \emph{dimensional} energy is the sum of the surface, $E_S$, and gravitational energy, $E_V$. Initially, we have only fluid $B$ in the vessel, so that
\begin{equation}
E_0 =E_{S,0} + E_{V,0}=  \sigma_3 \pi {\cal R}_0^2 R_{wall}^2 +  \frac{\pi}{2} \rho_B g  {\cal R}_0^4 h_f^2 R_{wall}^2,
\label{eq:Energy0}
\end{equation}
since we neglect the gravitational energy of fluid $C$ (air). In order to calculate the final energy, 
\begin{equation}
E = E_S + E_V,
\end{equation}
\label{eq:TotEnerg}
we consider the surface contributions of the three interfaces plus the gravitational contributions for liquids $A$ and $B$:

\begin{subequations}
\begin{equation}
E_S = 2 \pi {\cal R}_0^2 \left(\sigma_1 L_1 \int_0^1 r_1(q) dq + \sigma_2 L_2 \int_0^1 r_2(q) dq + \sigma_3 L_3 \int_0^1 r_3(q) dq \right),
\label{eq:SufEnerg}
\end{equation}
\begin{equation}
\begin{aligned}
E_V =\, &\pi \rho_A g {\cal R}_0^4 \left( \int_0^1 (z_1^2(q)-z_{tp}^2) r_1(q) r_1'(q) dq + \int_0^1 (z_{tp}^2-z_2^2(q)) r_2(q) r_2'(q) dq \right) +\\
& \pi \rho_B g {\cal R}_0^4 \left( \int_0^1 z_2^2(q) r_2(q) r_2'(q) dq + \int_0^1 z_3 r_3(q) r_3'(q) dq \right),
\label{eq:VolEnerg}
\end{aligned}
\end{equation}
\end{subequations}
where the integrals inside the parentheses are dimensionless quantities. Here, $z_{tp}$ is the $z$--coordinate of the triple point and $z=0$ corresponds to the bottom of the control region. The thickness $h_f$ is large enough to assure that the drops do not touch the solid substrate for all of the calculations.  

\begin{figure}[htb]
\subfigure[]{\includegraphics[width=0.3\linewidth]{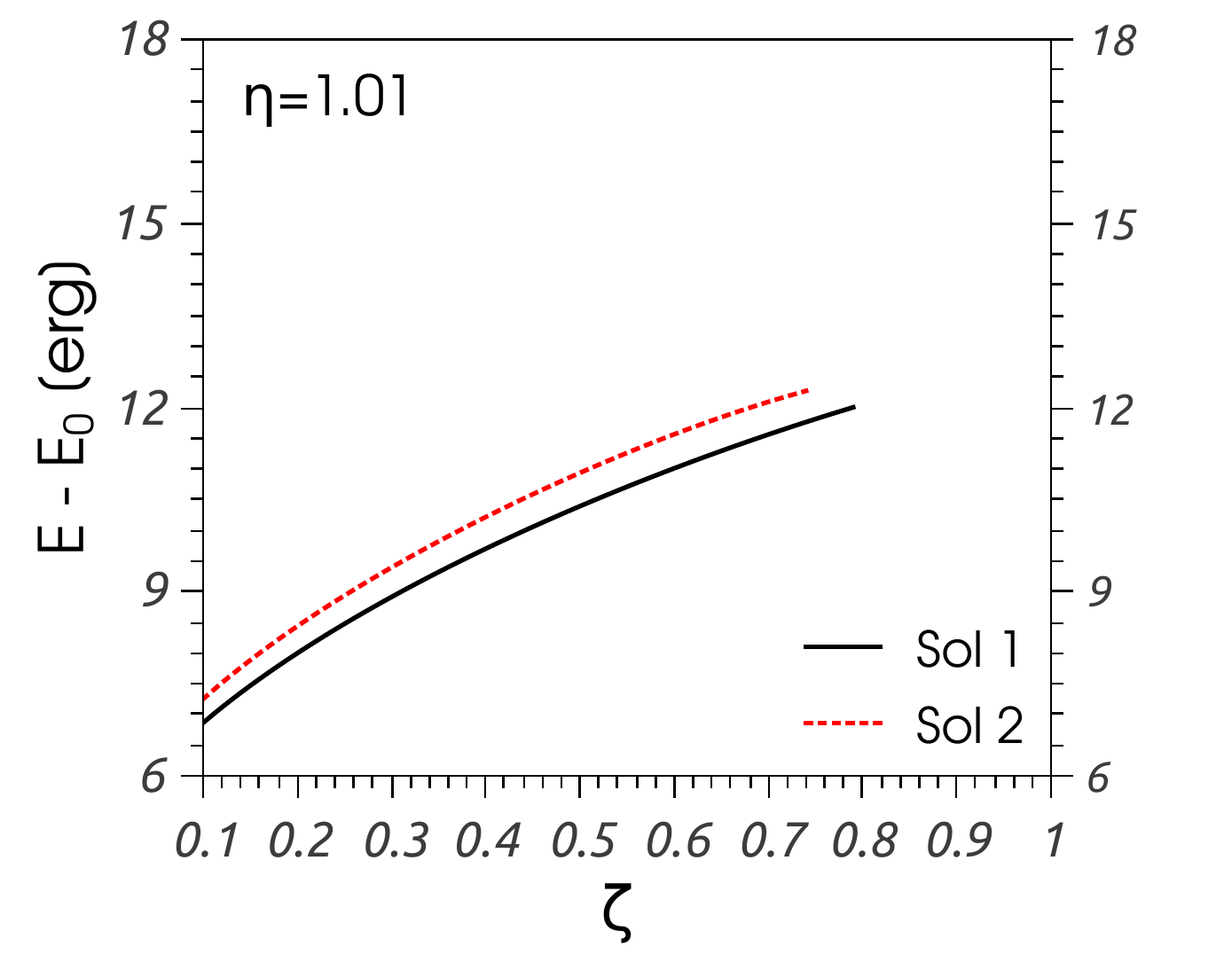}}
\subfigure[]{\includegraphics[width=0.3\linewidth]{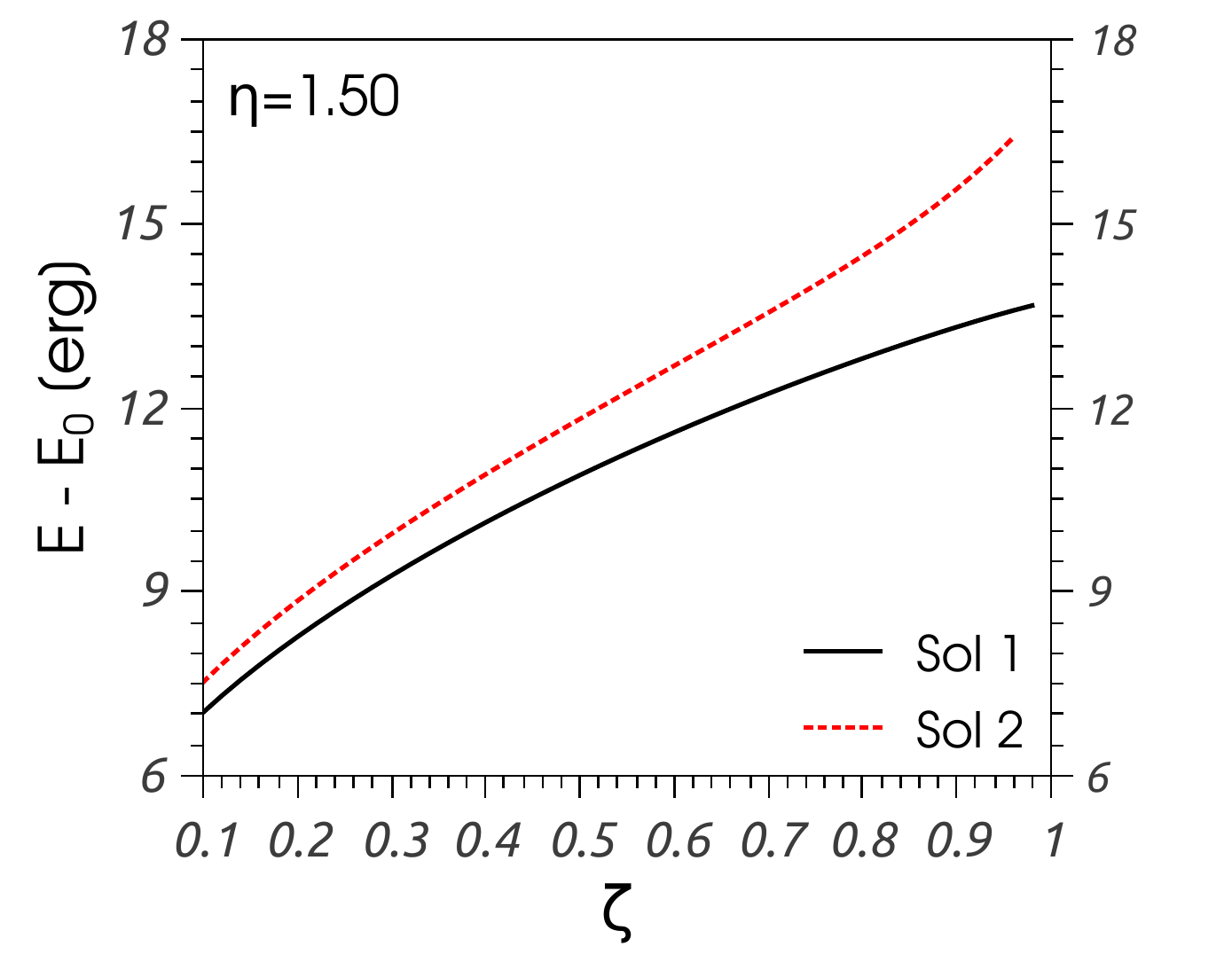}}
\subfigure[]{\includegraphics[width=0.3\linewidth]{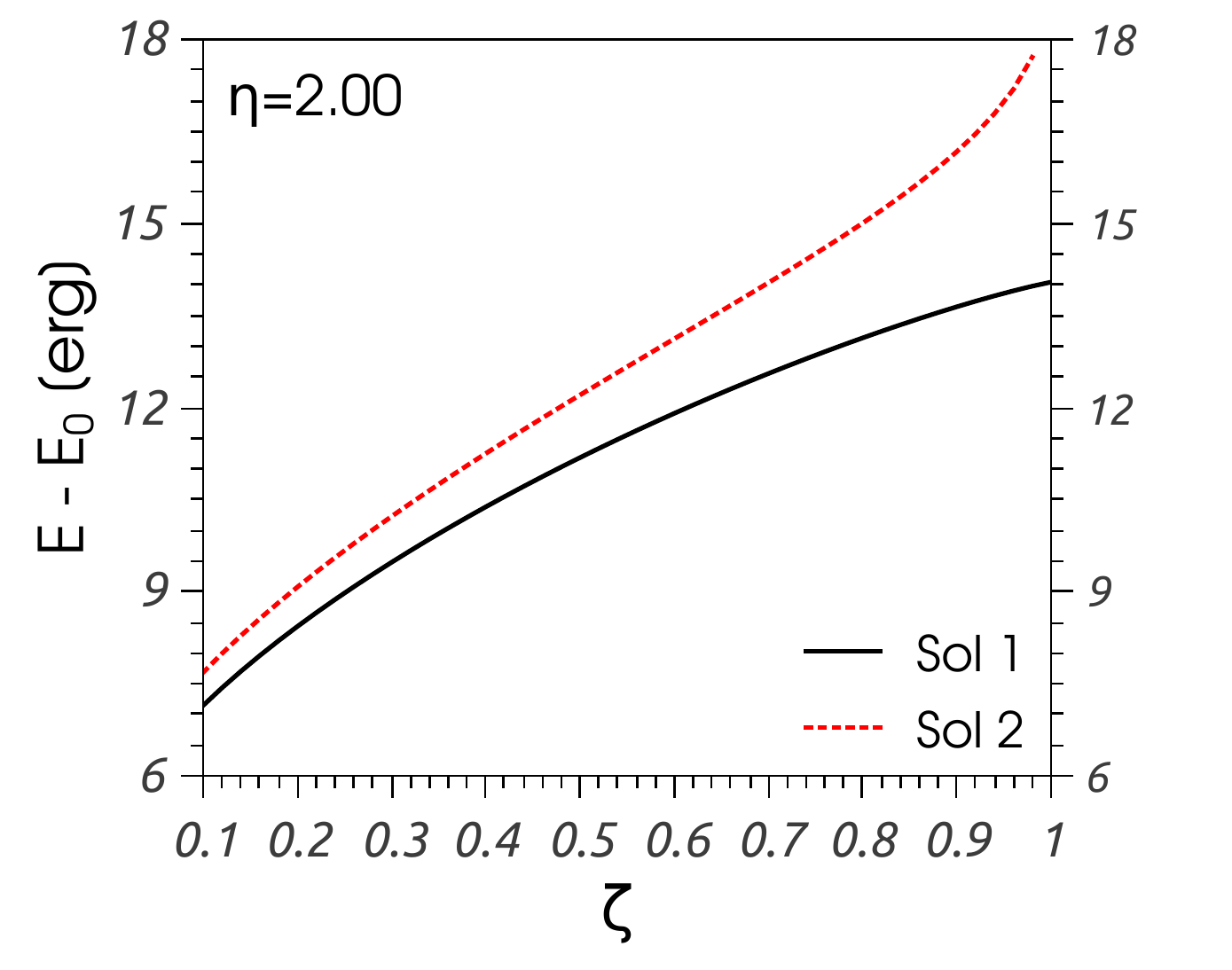}}
\caption{Case~A: Total energy variation for $Bo=1.22$, Sol~$1$ (solid lines) and Sol~$2$ (dashed lines). (a) $\eta=1.01$, (b) $\eta=1.5$, (c) $\eta=2$. The curves in (a) do not go beyond $\zeta \approx 0.8$ because of precision issues in the numerical scheme appearing when both $\xi$ and $\eta$ are very close to unity (their respective maximum and minimum limiting values).}
\label{fig:totalenergy}
\end{figure}

The variation of the total energy for the two families of solutions is shown for case A in Fig.~\ref{fig:totalenergy}. We find that Sol~$1$ has always lower energy than Sol~$2$ for different combinations of $(\eta, \zeta)$. The energy difference between the two solutions increases as both $\eta$ and $\zeta$ increase. 

\begin{figure}[htb]
\subfigure[]{\includegraphics[width=0.3\linewidth]{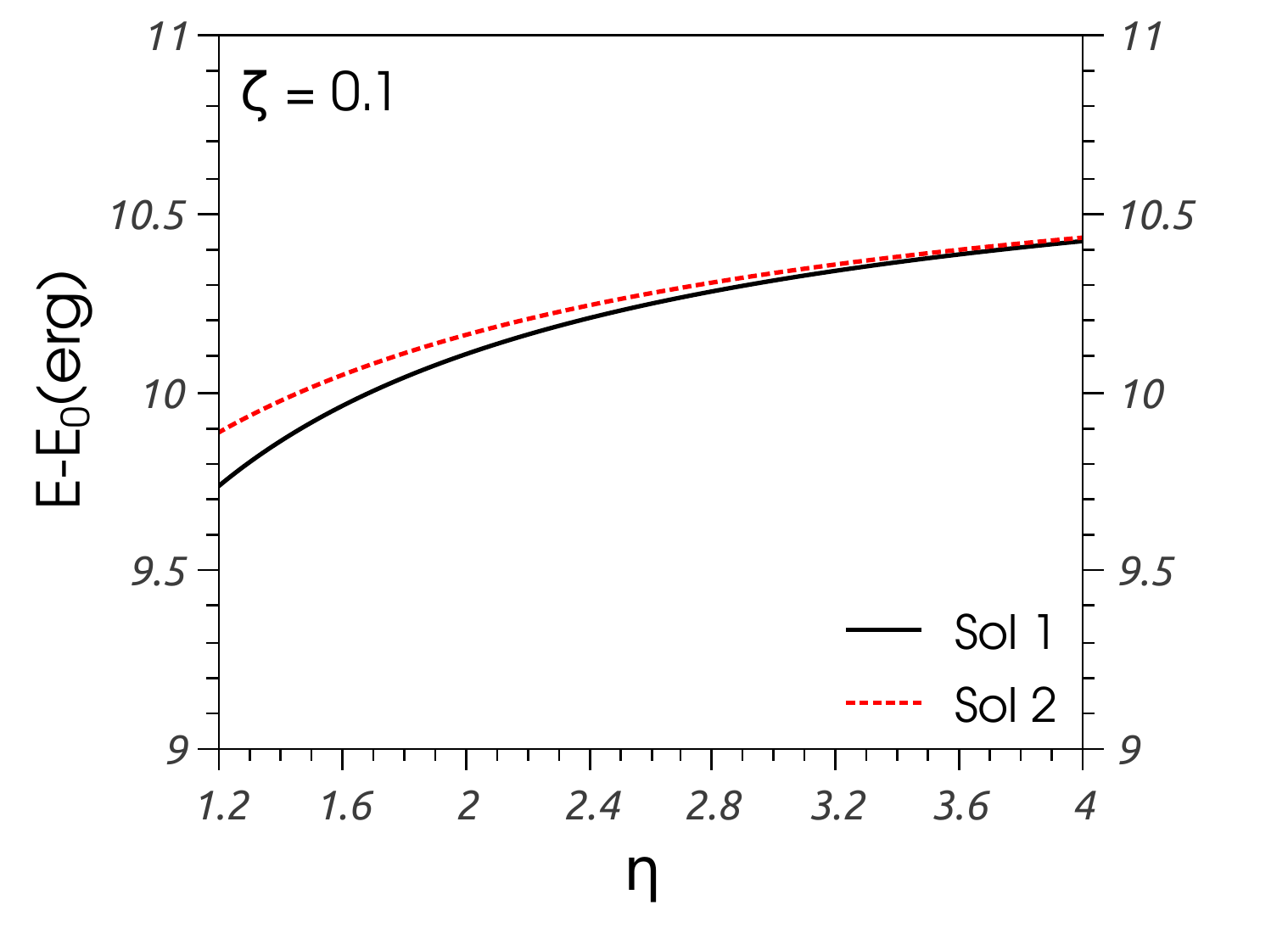}}
\subfigure[]{\includegraphics[width=0.3\linewidth]{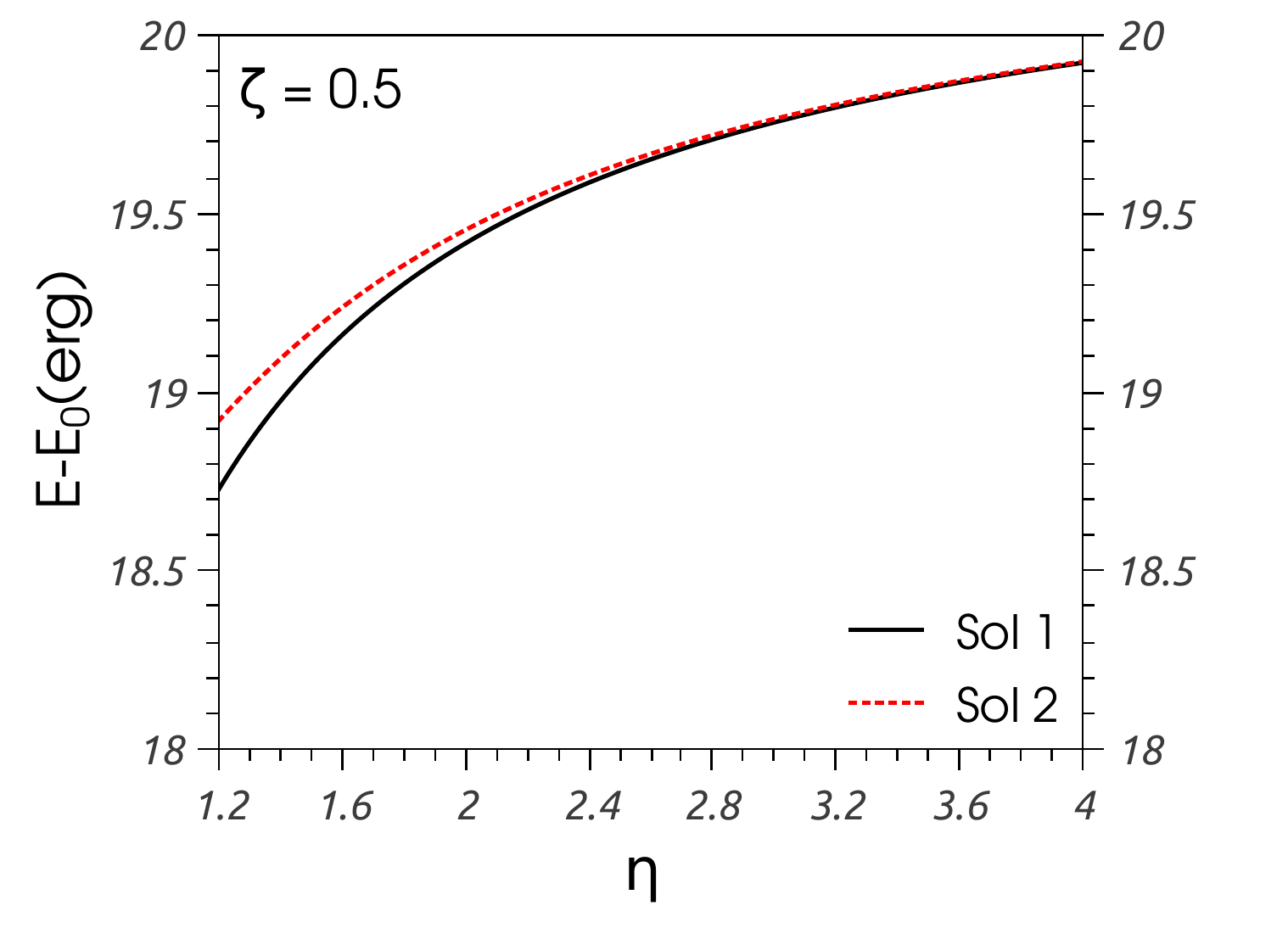}}
\subfigure[]{\includegraphics[width=0.3\linewidth]{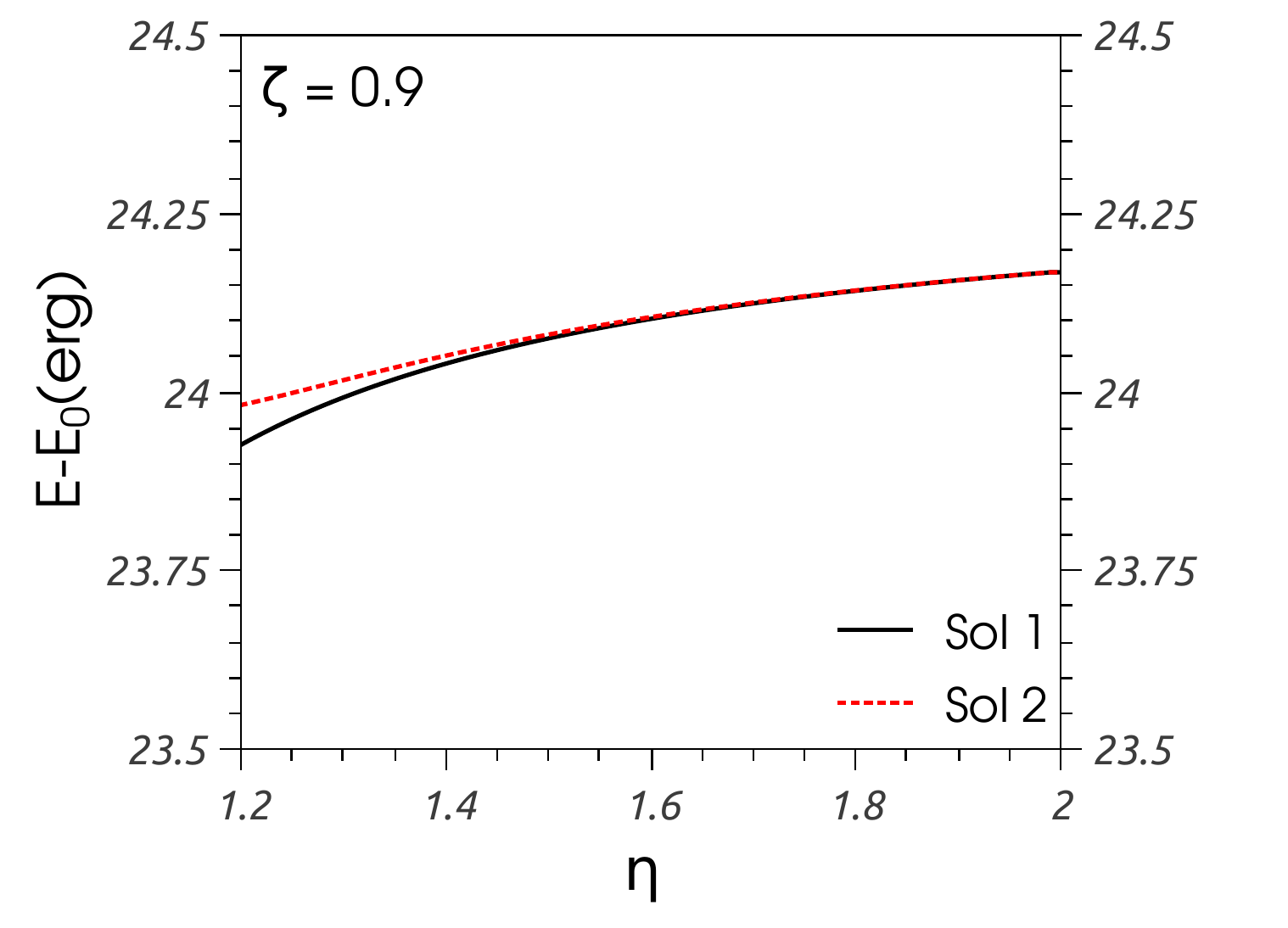}}
\caption{Case~B:  Total energy variation for $Bo=0.49$, Sol~$1$ (solid lines) and Sol~$2$ (dashed lines). (a) $\zeta=0.1$, (b) $\zeta=0.5$, (c) $\zeta=0.9$.}
\label{fig:totalenergyB}
\end{figure}

On the other hand, for case~B, we find the curves shown in Fig.~\ref{fig:totalenergyB}. In this case, Sol~$1$ also remains with lower energy than Sol~$2$ for $\eta \approx 1$, but their difference decreases as $\eta$ increases. This decrease occurs faster as $\zeta$ is closer to unity. This is consistent with the result that both solutions tend to converge to a single solution in this case. In summary, the energy analysis shows that Sol~$1$ is more likely to occur in nature for both cases A and B,  since it is always the lower energy.

\section{Summary and Conclusions}

Although there are many physical parameters involved in the determination of the shape of a liquid lens, we present here a dimensionless scheme that embraces all possible physical situations. The use of the parameters $\eta$ and $\zeta$ (see Eqs.~(\ref{eq:zetaS})--(\ref{eq:etas})) allows us to describe the behavior of the solutions for any combination of the three surface tensions $\sigma_1,\, \sigma_2$ and $\sigma_3$.

Within this framework, we have analytically solved the case without gravity and numerically analyzed the case with gravity. The no--gravity solution is presented as a tool to obtain the initial guesses for the numerical analysis performed in the more general case with gravity. A remarkable fact, not usually mentioned in the literature, is the requirement that the spreading factor $S$ is bounded from below, i.e., $S^\ast<S<0$, for equilibrium solutions to exist. This implies $0<\zeta<1$, which is more restrictive than the usual case for liquid drops on solids.
 
We have highlighted that the effects of gravity on the solution are far from trivial. For example, it is possible to obtain two different families of solutions for the same set of physical parameters. Although some of these solutions have been reported in the literature, this issue has apparently been unrecognized because the authors have assumed that, no matter is the set of guess values, any converged solution is valid based on the belief that it is unique. However, we show here that the solutions can be non--unique, and hence, a more rigurous treatment is needed. As mentioned before, the proof of uniqueness of the solution for the mathematical problem posed by Eqs.~(\ref{eq:rizi})--(\ref{eq:lens_vol}) is out of the scope of the present work. The main difference between the solution families is the shape of the free surface of the liquid $B$ near the lens, where the curvature of curve $3$ (see Fig.~\ref{fig:2solComp}) adopts a different sign for each family. 

In order to decide which solution is more likely to be found in nature, we also perform an energy analysis to compare the two families of solutions under two possible scenarios, namely case~A ($\sigma_1 < \sigma_2$) or case~B ($\sigma_1 > \sigma_2$). This analysis is done considering a finite volume of liquid $B$ (so it is contained in a vessel). It turns out that in both scenarios Sol~$1$ is always less energetic than Sol~$2$, so that it is most likely to be found in natural situations. Moreover, it is found that the two solution families converge to a unique one when $\sigma_1 > \sigma_2$, and both $\eta$ and $\zeta$ are large enough. 

\acknowledgments
PDR, AGG and JAD acknowledge support from Consejo Nacional de Investigaciones Cient\'{\i}ficas y T\'ecnicas (\mbox{CONICET}, Argentina) and Agencia Nacional de Promoci\'on Cient\'{\i}fica y Tecnol\'ogica (ANPCyT, Argentina) with grant PICT~1067/2016. HAS research was partially supported by NSF through the Princeton University’s Materials Research Science and Engineering Center DMR-1420541. 

\newpage
\appendix
\section{Initial configuration of the numerical procedure}\label{ApendixGuess}

As mentioned in Section~\ref{sec:GravitySolutions}, three possibilities arise from the choice of the guess  values in the numerical scheme: no convergence, or convergence to Sol~1 or Sol~2 families. 
These two kinds of solutions were found before (see Fig.~4 in~\cite{Burton2010}), but with different sets of parameters. So, one set led to a solution with concave curvature of surface $3$ and the other one with convex curvature. Thus, it was not possible to assert that they belonged to different families, because of the belief that the solution is unique. Therefore, there was no mention on the possibility of finding them for the same set of physical parameters. 

To clarify this issue, we perform a detailed analysis on the selection of the guess values for the iteration procedure for the case with $\rho_A=0.7\, g/cm^3$, $\rho_B=1.0\, g/cm^3$, $\sigma_1 = 25\, mN/m$, $\sigma_2 = 55\, mN/m$, $\sigma_3 = 70\, mN/m$ and ${\cal V}_0=0.579\,cm^3$, as presented in Fig.~4(a) in~\cite{Burton2010}. 

\begin{figure}[htb]
\includegraphics[width=0.6\linewidth]{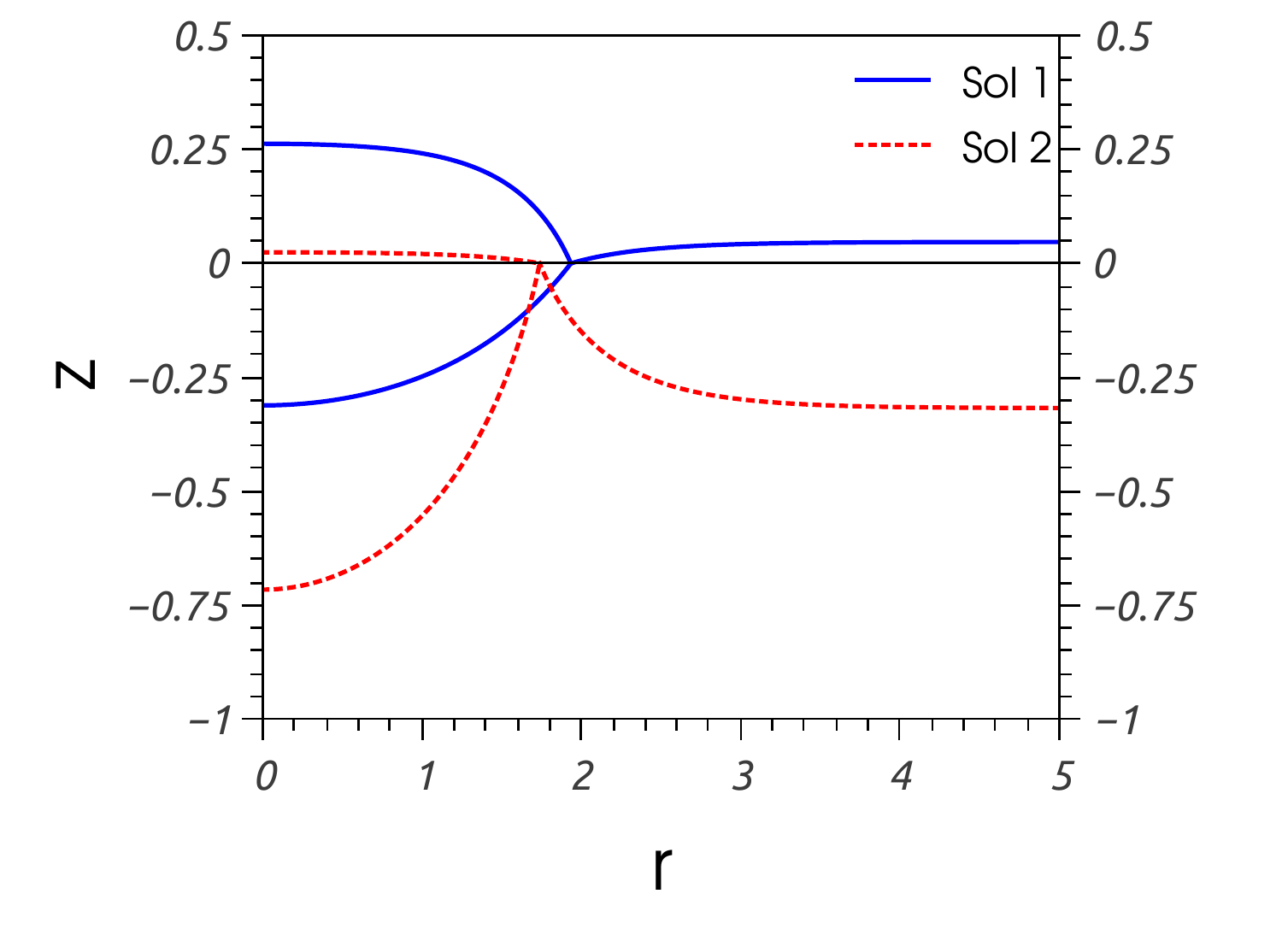}
\caption{Sol~$1$ (blue, solid) reported in Fig.~4(a) in~\cite{Burton2010}, and Sol~$2$ (red, dashed) the other possible solution obtained for the same set of physical parameters (see text).}
\label{fig:PerfBurt}
\end{figure}

Firstly, we calculate the corresponding dimensionless parameters $\eta=2.2$ and $\zeta=0.2$, as well as the length scale given by Eq.~(\ref{eq:R0}). Then, we follow the procedure described in Section~\ref{sec:analyticalSol} to analytically obtain the solution without gravity. Within this framework, we find $R_d=1.64811$, $L_1=1.81602$, $L_2=1.66852$, $R_1=2.35493$, $R_2=5.18084$, $\alpha=44.4^{\circ}$ and $\beta=18.5^{\circ}$. With these results and $L_3^{(0)}=R_{wall}-R_d^{(0)}$, we can construct two sets of guess values for $G=(\Delta P_1^{(0)}, \Delta P_2^{(0)}, \Delta P_3^{(0)}, L_1^{(0)}, L_2^{(0)}, L_3^{(0)}, R_d^{(0)})$, namely
\begin{eqnarray}
G_1&=&(-0.2,0.5,+0.0001,1.81602,1.66852,4.15407,1.64811), \\
G_2&=&(-0.2,0.5,-0.0001,1.81602,1.66852,4.15407,1.64811),
\end{eqnarray}
where $\Delta P_1^{(0)}$ and $\Delta P_2^{(0)}$ were chosen order one with different sign between each other, and $\Delta P_3^{(0)}\sim 10^{-4}$ has different sign in each set of parameters. This slightly difference in the initial values leads to the two solution families. The surface profiles obtained with these sets of guess values are shown in Fig.~\ref{fig:PerfBurt}. The resulting values for the solution sets $T=(\Delta P_1, \Delta P_2, \Delta P_3, L_1, L_2, L_3, R_d)$ are:
\begin{eqnarray}
T_1&=&(-0.05404,0.23077,+0.00012,1.97932,1.97290,3.86900,1.93438)\\
T_2&=&(-0.00888,0.57527,-0.00049,1.73873,1.95947,4.12058,1.73836).
\label{eq:BurtRsts}
\end{eqnarray}
According to our analysis in Section~\ref{sec:energy}, the Sol~$2$ family corresponds to a higher energy case, so it is not probable to be found in nature. Fortunately, Sol~$1$ was correctly reported in~\cite{Burton2010}.

\newpage
\bibliographystyle{unsrt}
\bibliography{liqiuidlenses.bib}

\end{document}